\documentclass[preprints,article,accept,oneauthor,pdftex]{Definitions/mdpi}

\firstpage{1}
\pubvolume{xx}
\issuenum{1}
\articlenumber{5}
\pubyear{2019}
\copyrightyear{2019}
\history{Received: date; Accepted: date; Published: date}

\Title{Probability Density Functions in Homogeneous and Isotropic Magneto-Hydrodynamic Turbulence}

\Author{Jan Friedrich $^{1,*}$}

\AuthorNames{Jan Friedrich}

\address{%
$^{1}$ \quad Univ. Lyon, ENS de Lyon, Univ. Claude Bernard, CNRS, Laboratoire de Physique,
F-69342, Lyon, France; jan-friedrich1@ens-lyon.fr}

\corres{Correspondence: jan-friedrich1@ens-lyon.fr}

\abstract{We derive a hierarchy of evolution equations for multi-point probability density functions in magneto-hydrodynamic (MHD) turbulence. We discuss the relation to the moment hierarchy in MHD turbulence derived by Chandrasekhar [S. Chandrasekhar, Proc. R. Soc. Lond. A 1951, 204, 435–449]
and derive a functional equation for a joint characteristic functional which can be considered as the analogon to the Hopf functional in hydrodynamic turbulence.
Furthermore, we develop a closure method for the evolution equation of the single-point magnetic field probability density function which is based on a joint Gaussian assumption for unclosed terms. It is explicitly shown that this closure, together with the assumptions of homogeneity and isotropy, leads to vanishing nonlinear terms. We discuss the implications
of this finding for magnetic field generation and give a brief outlook on an axisymmetric theory which includes a mean magnetic field.}


\begin{document}

\section{Introduction}
\label{sec:intro}
The investigation of magnetohydrodynamic (MHD) turbulence by statistical methods
has a longstanding tradition which can be traced back to the works of Chandrasekhar~\cite{chandra:1951} and Batchelor~\cite{Batchelor1950}, as well as to the subdivision of mean field electrodynamics put forth by Steenbeck, Krause and R\"adler~\cite{steenbeck} (for further references, see~\cite{Parker1996,Radler2007}). Recent developments in statistical magnetohydrodynamics include an analytical treatment of weak MHD turbulence~\cite{galtier_2000}, a phenomenological description
for the energy spectrum based on the dynamical alignment of velocity and
magnetic field fluctuations~\cite{Boldyrev2006,Perez2010}, and the derivation of relations between longitudinal and transverse structure functions~\cite{Friedrich2016} similar to the ones of hydrodynamic turbulence
~\cite{hill:2001,yakhot:2006,grauer-homann-pinton:2012}. On the other hand, improved measurements in the solar wind~\cite{PhysRevLett.99.115001,Fraternale_2019} along with an ever-growing number of experiments, which operate with liquid sodium
under turbulent conditions in order to
study the dynamo effect~\cite{Bourgoin2002,giesecke_2018,shew2002mechanically}, are of great importance for the direct assessment of certain statistical quantities.

Therefore, a comprehensive statistical formulation of MHD turbulence will also lead to considerable advances in the modeling of such astrophysical and experimental MHD flows. Nonetheless, similar to the case of hydrodynamic turbulence, such a statistical description of MHD turbulence is complicated by the occurrence of anomalous statistics of velocity and magnetic field fluctuations at small scales which is commonly referred to as intermittency~\cite{frisch:1995}.
To some degree intermittency effects in fully developed MHD turbulence are even more pronounced than in ordinary turbulence, a fact that is commonly attributed to the geometrical structures observed in MHD turbulence, i.e., nearly singular current and vortex sheets~\cite{Grauer1996,politano1995}. Moreover, the Alfv\'en effect~\cite{chandra:1961}  introduces additional nonlocalities to the multi-point hierarchy of MHD flows which will be discussed in more detail in Sec.~\ref{sec:hierarchy}.
Therefore, due to pronounced deviations from Gaussianity and additional nonlocalities in the multi-point hierarchy of MHD turbulence, finding appropriate methods or assumptions to close the hierarchy at a certain stage~\cite{Friedrich2018,Friedrich2017}
might prove to be even more difficult than in ordinary turbulence.

However, the presence of the magnetic field in conducting turbulent flows can also have a regularizing effect on the nonlinearities of the MHD equations. For instance, as shown by direct numerical simulations of MHD turbulence~\cite{Friedrich2016}, the additional possibility for the velocity field to align with the  magnetic field leads to depleted pressure contributions in regions with preferred alignment (anti-alignment).

In the following we derive evolution equations for multi-point velocity and magnetic field probability density functions (PDFs) directly from the MHD equations. The emerging hierarchy can be considered as the counterpart to the hierarchy in hydrodynamic turbulence  by Lundgren~\cite{Lundgren1967}, Monin~\cite{Monin1967}, and Novikov~\cite{Novikov1968} (we also refer the reader to ~\cite{Friedrich2012a} for further reviews)
and is similar to the BBGKY hierarchy of statistical mechanics. We will then proceed to discuss a possible closure method on the basis of a joint Gaussian assumption for velocity and magnetic field statistics.
\section{Derivation of a Hierarchy of PDF equations}
\label{sec:hierarchy}
The purpose of this section is to derive evolution equations for multi-point velocity and magnetic field
PDFs in an analogous manner to the hydrodynamic case discussed in the seminal work by Lundgren~\cite{Lundgren1967}.
To this end, we consider the MHD equations in the following form
\begin{align}\label{eq:u}
  \frac{\partial}{\partial t}\mathbf{u}(\mathbf{x},t)+ \mathbf{u}(\mathbf{x},t) \cdot \nabla_{\mathbf{x}}\mathbf{u}(\mathbf{x},t)
  - \mathbf{h}(\mathbf{x},t) \cdot \nabla_{\mathbf{x}}\mathbf{h}(\mathbf{x},t)
  &= -\nabla_{\mathbf{x}}p(\mathbf{x},t) +\nu \nabla_{\mathbf{x}}^2 \mathbf{u}(\mathbf{x},t)\;,\\
  \frac{\partial}{\partial t}\mathbf{h}(\mathbf{x},t)+ \mathbf{u}(\mathbf{x},t) \cdot \nabla_{\mathbf{x}}\mathbf{h}(\mathbf{x},t)
  - \mathbf{u}(\mathbf{x},t) \cdot \nabla_{\mathbf{x}}\mathbf{h}(\mathbf{x},t)
  &= \lambda \nabla_{\mathbf{x}}^2 \mathbf{h}(\mathbf{x},t)\;,
  \label{eq:h}
\end{align}
where $\nu$ denotes the kinematic viscosity, $\lambda$ the magnetic diffusivity,
and $p(\mathbf{x},t)$ the total pressure, i.e., hydrodynamic and magnetic pressure. Furthermore, it should be stressed that in this particular form of the MHD equations, the magnetic field has the dimensions of a velocity.
The MHD equations are completed by the incompressibility conditions
$\nabla \cdot \mathbf{u}(\mathbf{x},t)$ and
$\nabla \cdot \mathbf{h}(\mathbf{x},t)$.\\
We define single-point magnetic and velocity field PDFs according to
\begin{align}\label{eq:f_u}
  f^{\mathbf{u}}(\mathbf{v}, \mathbf{x}, t)&=\langle \delta(\mathbf{v}-\mathbf{u}(\mathbf{x},t)) \rangle\;,\\
  f^{\mathbf{h}}(\mathbf{b}, \mathbf{x}, t)&=\langle \delta(\mathbf{b}-\mathbf{h}(\mathbf{x},t)) \rangle\;.
  \label{eq:f_h}
\end{align}
In principle, the following treatment could also be carried out in terms of Els\"asser fields $\mathbf{z}^{\pm}=\mathbf{u}\pm \mathbf{h}$. However,
due to the different transformation behavior of $\mathbf{h}$ (axial vector) and $\mathbf{u}$ (polar vector) under reflexions, it will be more appropriate
to work in the original fields,
especially with regard to the tensor calculus of isotropic and homogeneous MHD turbulence~\cite{Robertson1940,chandra:1951,Friedrich2016} that will be used in Sec.~\ref{sec:closure}.
For later convenience, we also define the two-point PDFs
\begin{align}\label{eq:f_uu}
  f^{\mathbf{u}\mathbf{u}}(\mathbf{v}, \mathbf{x};\mathbf{v'}, \mathbf{x'}, t)&=\langle\delta(\mathbf{v}-\mathbf{u}(\mathbf{x},t)) \delta(\mathbf{v'}-\mathbf{u}(\mathbf{x'},t)) \rangle\;,\\
  \label{eq:f_uh}
  f^{\mathbf{u}\mathbf{h}}(\mathbf{v}, \mathbf{x};\mathbf{b}, \mathbf{x'}, t)&=\langle\delta(\mathbf{v}-\mathbf{u}(\mathbf{x},t)) \delta(\mathbf{b}-\mathbf{h}(\mathbf{x'},t)) \rangle\;,
  \\
  f^{\mathbf{h}\mathbf{h}}(\mathbf{b}, \mathbf{x};\mathbf{b'}, \mathbf{x'}, t)&=\langle\delta(\mathbf{b}-\mathbf{h}(\mathbf{x},t)) \delta(\mathbf{b'}-\mathbf{h}(\mathbf{x'},t)) \rangle\;.
    \label{eq:f_hh}
\end{align}
The evolution equation for the single-point velocity field can be obtained by deriving the single-point velocity PDF (\ref{eq:f_u}) with respect to time
\begin{equation}
  \frac{\partial}{\partial t}   f^{\mathbf{u}}(\mathbf{v}, \mathbf{x}, t)
  = - \nabla_{\mathbf{v}} \cdot \left \langle \frac{\partial \mathbf{u}(\mathbf{x},t) }{\partial t}\delta(\mathbf{v}-\mathbf{u}(\mathbf{x},t)) \right \rangle\;.
  \label{eq:evo_u_init}
\end{equation}
Here, we have to insert the temporal evolution of the velocity field which is given by eq. (\ref{eq:u}) and treat the corresponding terms separately. The treatment is similar to the hydrodynamic case~\cite{Lundgren1967} and is discussed in Appendix~\ref{app:f_u}.
The evolution equation for the single-point velocity PDF thus reads
\begin{align}\nonumber
\lefteqn{ \left[ \frac{\partial}{\partial t} + \mathbf{v} \cdot \nabla_{\mathbf{x}}\right]
f^{\mathbf{u}}(\mathbf{v}, \mathbf{x}, t) }\\
 \nonumber
 = &-\nabla_{\mathbf{v}} \cdot  \int \textrm{d}\mathbf{x'}\delta(\mathbf{x}-\mathbf{x'}) \int \textrm{d}\mathbf{b}\; \left(\mathbf{b}\cdot \nabla_{\mathbf{x'}}\right)
 \mathbf{b}  f^{\mathbf{u}\mathbf{h}}(\mathbf{v}, \mathbf{x}; \mathbf{b}, \mathbf{x'}, t)
 \\ \nonumber
 &+\frac{1}{4\pi} \nabla_{\mathbf{v}}\cdot
 \int \textrm{d}\mathbf{x'} \left[\nabla_{\mathbf{x}} \frac{1}{|\mathbf{x}-\mathbf{x'}|}\right] \int \textrm{d}\mathbf{v'} \left(\mathbf{v'}\cdot \nabla_{\mathbf{x'}} \right)^2
 f^{\mathbf{u}\mathbf{u}}(\mathbf{v}, \mathbf{x}; \mathbf{v'}, \mathbf{x'}, t)\\ \nonumber
 & -\frac{1}{4\pi} \nabla_{\mathbf{v}}\cdot
 \int \textrm{d}\mathbf{x'} \left[\nabla_{\mathbf{x}} \frac{1}{|\mathbf{x}-\mathbf{x'}|}\right] \int \textrm{d}\mathbf{b} \left(\mathbf{b}\cdot \nabla_{\mathbf{x'}} \right)^2
 f^{\mathbf{u}\mathbf{h}}(\mathbf{v}, \mathbf{x}; \mathbf{b}, \mathbf{x'}, t)\\
 &-\nu \nabla_{\mathbf{v}} \cdot \int \textrm{d}\mathbf{x'} \delta(\mathbf{x}-\mathbf{x'}) \nabla_{\mathbf{x'}}^2 \int \textrm{d}\mathbf{v'} \mathbf{v'}
 f^{\mathbf{u}\mathbf{u}}(\mathbf{v}, \mathbf{x}; \mathbf{v'}, \mathbf{x'}, t)\;,
\label{eq:evo_u}
\end{align}
and therefore couples to the two-point PDFs (\ref{eq:f_uu}-\ref{eq:f_uh}).
By the same token we can derive the single-point magnetic field PDF (\ref{eq:f_h})
with respect to time
\begin{equation}
  \frac{\partial}{\partial t}   f^{\mathbf{h}}(\mathbf{b}, \mathbf{x}, t)
  = - \nabla_{\mathbf{b}} \cdot \left \langle \frac{\partial \mathbf{h}(\mathbf{x},t) }{\partial t}\delta(\mathbf{b}-\mathbf{h}(\mathbf{x},t)) \right \rangle\;,
  \label{eq:evo_h_init}
\end{equation}
insert the evolution equation (\ref{eq:h}), and relate the corresponding terms to the two-point PDFs (\ref{eq:f_uh}-\ref{eq:f_hh}), which is discussed in Appendix~\ref{app:f_h}.
The counterpart to the single-point velocity PDF equation (\ref{eq:evo_u}), i.e., the evolution equation
for the single-point magnetic field PDF thus reads
\begin{align}\nonumber
 \lefteqn{\frac{\partial}{\partial t}   f^{\mathbf{h}}(\mathbf{b}, \mathbf{x}, t)
 +\int \textrm{d}\mathbf{v} \mathbf{v}\cdot \nabla_{\mathbf{x}}
   f^{\mathbf{u}\mathbf{h}}(\mathbf{v}, \mathbf{x};\mathbf{b}, \mathbf{x}, t)}\\
   \nonumber
   =& -\nabla_{\mathbf{b}} \cdot \int \textrm{d}\mathbf{x'}\delta(\mathbf{x}-\mathbf{x'})  \int \textrm{d}\mathbf{v} \left(\mathbf{b}\cdot \nabla_\mathbf{x'} \right) \mathbf{v}
   f^{\mathbf{u}\mathbf{h}}(\mathbf{v}, \mathbf{x'};\mathbf{b}, \mathbf{x}, t)\\
   & -\lambda \nabla_{\mathbf{b}} \cdot \int \textrm{d}\mathbf{x'} \delta(\mathbf{x}-\mathbf{x'}) \nabla_{\mathbf{x'}}^2 \int \textrm{d}\mathbf{b'} \mathbf{b'}
   f^{\mathbf{h}\mathbf{h}}(\mathbf{b}, \mathbf{x}; \mathbf{b'}, \mathbf{x'}, t)\;,
   \label{eq:evo_h}
\end{align}
which, again, is an unclosed equation since it couples to
the two-point quantities (\ref{eq:f_uh}-\ref{eq:f_hh}). Eqs. (\ref{eq:evo_u}) and (\ref{eq:evo_h}) are the first equations in an infinite chain of evolution equations, e.g., the two-point PDFs (\ref{eq:f_uu}-\ref{eq:f_hh}) will couple to three-point PDFs and so forth. Typical assumptions that are used in order to close such hierarchies include the assumption of statistical independence as well as Gaussian assumptions~\cite{Friedrich2012a}
and will be further discussed in Sec.~\ref{sec:gauss}.

In comparison to ordinary turbulence where unclosed terms are due to nonlocal pressure contributions and the viscous term, the Alfv\'en effect introduces additional nonlocalities
to the multi-point hierarchy. Indeed, it can readily be seen that both the l.h.s. of eq. (\ref{eq:evo_u}) and eq.  (\ref{eq:evo_h}) are Galilei invariant , i.e.,  they are unchanged under the transformation $\mathbf{v} \rightarrow \mathbf{v}+\mathbf{V}$, $\mathbf{b} \rightarrow \mathbf{b}$, and $\mathbf{x}
\rightarrow \mathbf{x}-\mathbf{V}t$, whereas for $\mathbf{v} \rightarrow \mathbf{v}$ and $\mathbf{b} \rightarrow \mathbf{b}+\mathbf{B}$, $\mathbf{B}$ cannot be removed by a Galilei transformation due to the first terms on the r.h.s of eq. (\ref{eq:f_u}) and eq.  (\ref{eq:f_h}). Obviously, latter transformation behavior
is inherited from the basic MHD equations (\ref{eq:u}-\ref{eq:h}).

In order to generalize the results for the one-point magnetic and velocity field PDF to an arbitrary number of points, we define the
 $(n+m)$-point joint velocity-magnetic field PDF
\begin{equation}
  f^{n\mathbf{u},m\mathbf{h}}(\{\mathbf{v}_i, \mathbf{x}_i\};\{\mathbf{b}_j, \mathbf{y}_j\},t)
  = \prod_{i=1}^n \prod_{j=1}^m \left \langle \delta(\mathbf{v}_i- \mathbf{u}(\mathbf{x}_i),t)
  \delta(\mathbf{b}_j- \mathbf{h}(\mathbf{y}_j),t) \right \rangle\;.
  \label{eq:multi_pdf}
\end{equation}
A treatment similar to the one described in Appendix~\ref{app:deriv_hierarchy} yields the following evolution equation for the  $(n+m)$-point joint velocity-magnetic field PDF
\begin{align}\nonumber
\lefteqn{ \left[\frac{\partial}{\partial t}
 +\sum_{i=1}^n \mathbf{v}_i \cdot \nabla_{\mathbf{x}_i}
 \right]f^{n\mathbf{u},m\mathbf{h}}(\{\mathbf{v}_i, \mathbf{x}_i\},\{\mathbf{b}_j, \mathbf{y}_j\},t)}\\ \nonumber
&+ \sum_{j=1}^m\int \textrm{d} \mathbf{v}
 \mathbf{v} \cdot \nabla_{\mathbf{y}_j} f^{(n+1)\mathbf{u},m\mathbf{h}}(\mathbf{v}, \mathbf{y}_j;\{\mathbf{v}_i, \mathbf{x}_i\};\{\mathbf{b}_j, \mathbf{y}_j\},t)\\ \nonumber
=&-\sum_{i=1}^n \nabla_{\mathbf{v}_i} \cdot  \int \textrm{d}\mathbf{x}\delta(\mathbf{x}_i-\mathbf{x}) \int \textrm{d}\mathbf{b}\; \left(\mathbf{b}\cdot \nabla_{\mathbf{x}} \right)
\mathbf{b}   f^{n\mathbf{u},(m+1)\mathbf{h}}(\{\mathbf{v}_i, \mathbf{x}_i\};\mathbf{b},\mathbf{x};\{\mathbf{b}_j, \mathbf{y}_j\},t)\\ \nonumber
&-\sum_{j=1}^m \nabla_{\mathbf{b}_j} \cdot \int \textrm{d}\mathbf{y}\delta(\mathbf{y}_j-\mathbf{y})  \int \textrm{d}\mathbf{v} \left(\mathbf{b}_j\cdot
\nabla_{\mathbf{y}}\right) \mathbf{v}
f^{(n+1)\mathbf{u}, m\mathbf{h}}(\mathbf{v}, \mathbf{y};\{\mathbf{v}_i, \mathbf{x}_i\};\{\mathbf{b}_j, \mathbf{y}_j\},t)\\ \nonumber
&+\frac{1}{4\pi} \sum_{i=1}^n \nabla_{\mathbf{v}_i}\cdot
\int \textrm{d}\mathbf{x} \left[\nabla_{\mathbf{x}_i} \frac{1}{|\mathbf{x}_i-\mathbf{x}|}\right] \int \textrm{d}\mathbf{v} \left(\mathbf{v}\cdot \nabla_{\mathbf{x}} \right)^2
f^{(n+1)\mathbf{u},m\mathbf{h}}(\mathbf{v}, \mathbf{x};\{\mathbf{v}_i, \mathbf{x}_i\};\{\mathbf{b}_j, \mathbf{y}_j\},t)\\ \nonumber
&-\frac{1}{4\pi} \sum_{i=1}^n \nabla_{\mathbf{v}_i}\cdot
\int \textrm{d}\mathbf{x} \left[\nabla_{\mathbf{x}_i} \frac{1}{|\mathbf{x}_i-\mathbf{x}|}\right] \int \textrm{d}\mathbf{b} \left(\mathbf{b}\cdot \nabla_{\mathbf{x}} \right)^2
f^{n\mathbf{u},(m+1)\mathbf{h}}(\{\mathbf{v}_i, \mathbf{x}_i\};\mathbf{b},\mathbf{x};\{\mathbf{b}_j, \mathbf{y}_j\},t) \\ \nonumber
&-\nu \sum_{i=1}^n\nabla_{\mathbf{v}_i} \cdot \int \textrm{d}\mathbf{x} \delta(\mathbf{x}_i-\mathbf{x}) \nabla_{\mathbf{x}}^2 \int \textrm{d}\mathbf{v} \mathbf{v}
f^{(n+1)\mathbf{u},m\mathbf{h}}(\mathbf{v},\mathbf{x};\{\mathbf{v}_i, \mathbf{x}_i\};\{\mathbf{b}_j, \mathbf{y}_j\},t)\\
&-\lambda \sum_{j=1}^m\nabla_{\mathbf{b}_j} \cdot \int \textrm{d}\mathbf{y} \delta(\mathbf{y}_j-\mathbf{y}) \nabla_{\mathbf{y}}^2 \int \textrm{d}\mathbf{b} \mathbf{b}
f^{n\mathbf{u},(m+1)\mathbf{h}}(\{\mathbf{v}_i, \mathbf{x}_i\};\mathbf{b},\mathbf{y};\{\mathbf{b}_j, \mathbf{y}_j\},t)\;.
\label{eq:generalized}
\end{align}
Eq. (\ref{eq:generalized}) is the most general evolution equation for multi-point PDFs in MHD turbulence. In the following section, we will discuss a more compact statistical description of MHD turbulence in terms of functionals.
Here, we briefly want to relate the PDF hierarchy (\ref{eq:generalized}) to the moment hierarchy in MHD turbulence at the example of two-point correlation functions. To this end, we take the moments of the two-point PDFs (\ref{eq:f_uu}-\ref{eq:f_hh}) which results in the well-known two-point correlation functions of MHD turbulence
\begin{align}\label{eq:cuu}
  C^{\mathbf{u}\mathbf{u}}_{i\;j}(\mathbf{x};\mathbf{x'},t)
  =& \int \textrm{d} \mathbf{v}v_i \int \textrm{d}\mathbf{v'}v'_j f^{\mathbf{u}\mathbf{u}}(\mathbf{v}, \mathbf{x};\mathbf{v'}, \mathbf{x'}, t)=\left \langle u_i(\mathbf{x},t)
  u_j(\mathbf{x'},t) \right \rangle\;,\\
  \label{eq:cuh}
  C^{\mathbf{u}\mathbf{h}}_{i\;j}(\mathbf{x};\mathbf{x'},t)
  =& \int \textrm{d} \mathbf{v}v_i \int \textrm{d}\mathbf{b}b_j f^{\mathbf{u}\mathbf{h}}(\mathbf{v}, \mathbf{x};\mathbf{b}, \mathbf{x'}, t)= \left \langle u_i(\mathbf{x},t)
  h_j(\mathbf{x'},t) \right \rangle\;,\\
  C^{\mathbf{h}\mathbf{h}}_{i\;j}(\mathbf{x};\mathbf{x'},t)
  =& \int \textrm{d} \mathbf{b}b_i \int \textrm{d}\mathbf{b'}b'_j f^{\mathbf{h}\mathbf{h}}(\mathbf{b}, \mathbf{x};\mathbf{b'}, \mathbf{x'}, t)= \left \langle h_i(\mathbf{x},t)
  h_j(\mathbf{x'},t) \right \rangle\;,
  \label{eq:chh}
\end{align}
where the indices $i$ and $j$ now denote vector components.
It can be shown (see Appendix~\ref{app:chandra}) that in taking the moments (\ref{eq:cuu}-\ref{eq:chh}) of the evolution eqs. (\ref{eq:generalized}) for (\ref{eq:f_uu}-\ref{eq:f_hh}) one indeed obtains equations relating moments of second order to moments of third order. This moment hierarchy can be considered as the MHD analogon to the Friedmann-Keller hierarchy~\cite{Keller1924} and was
first derived by  Chandrasekhar~\cite{chandra:1951}  who also addressed the implications of homogeneity and isotropy in an invariant theory for MHD turbulence. We will make use of this invariant theory in Sec.~\ref{sec:closure} which discusses a closure method for the single-point magnetic field
PDF equation (\ref{eq:evo_h}). We will end this section with some general comments on the multi-point PDF (\ref{eq:multi_pdf}). First, fusing different points of the velocity and the magnetic field $\mathbf{y}_l \rightarrow \mathbf{x}_k$ in the evolution equation (\ref{eq:generalized}) can be achieved by the integration
\begin{align}\nonumber
    \lefteqn{f^{n\mathbf{u},m\mathbf{h}}(\mathbf{v}_1,\mathbf{x}_1;\ldots ;\mathbf{v}_k,\mathbf{x}_k; \ldots,\mathbf{v}_n,\mathbf{x}_n; \mathbf{b}_1,\mathbf{y}_1;\ldots \mathbf{b}_l,\mathbf{x}_k; \ldots;\mathbf{b}_m,\mathbf{y}_m,t)}\\
    &=
    \int \textrm{d}\mathbf{y}_l \delta(\mathbf{y}_l-\mathbf{x}_k) f^{n\mathbf{u},m\mathbf{h}}(\{\mathbf{v}_i, \mathbf{x}_i\};\{\mathbf{b}_j, \mathbf{y}_j\},t)\;.
\end{align}
This integration can be carried out directly for all terms in eq. (\ref{eq:generalized}) except for the third term for $j=l$ of the l.h.s which yields
\begin{align}\nonumber
  \lefteqn{\int \textrm{d}\mathbf{y}_l \delta(\mathbf{y}_l-\mathbf{x}_k) \int \textrm{d}\mathbf{v}\mathbf{v}
  \cdot \nabla_{\mathbf{y}_l}
  f^{(n+1)\mathbf{u},m\mathbf{h}}(\mathbf{v}, \mathbf{y}_l;\{\mathbf{v}_i, \mathbf{x}_i\};\{\mathbf{b}_j, \mathbf{y}_j\},t)}\\ \nonumber
  &=- \int \textrm{d}\mathbf{y}_l  (\underbrace{\nabla_{\mathbf{y}_l}\delta(\mathbf{y}_l-\mathbf{x}_k)}_{-\nabla_{\mathbf{x}_k}\delta(\mathbf{y}_l-\mathbf{x}_k)}) \cdot \int \textrm{d}\mathbf{v}\mathbf{v}
  f^{(n+1)\mathbf{u},m\mathbf{h}}(\mathbf{v}, \mathbf{y}_l;\{\mathbf{v}_i, \mathbf{x}_i\};\{\mathbf{b}_j, \mathbf{y}_j\},t)\\ \nonumber
  &= \int \textrm{d}\mathbf{v}\mathbf{v}\cdot \nabla_{\mathbf{x}_k} \int \textrm{d}\mathbf{y}_l  \delta(\mathbf{y}_l-\mathbf{x}_k)
  f^{(n+1)\mathbf{u},m\mathbf{h}}(\mathbf{v}, \mathbf{y}_l;\{\mathbf{v}_i, \mathbf{x}_i\};\{\mathbf{b}_j, \mathbf{y}_j\},t)\\ \nonumber
  &= \int \textrm{d}\mathbf{v}\mathbf{v}\cdot \nabla_{\mathbf{x}_k} \underbrace{f^{(n+1)\mathbf{u},m\mathbf{h}}(\mathbf{v},
  \mathbf{x}_k;\mathbf{v}_1,\mathbf{x}_1;\ldots ;\mathbf{v}_k,\mathbf{x}_k; \ldots;\mathbf{v}_n,\mathbf{x}_n; \mathbf{b}_1,\mathbf{y}_1;\ldots; \mathbf{b}_l,\mathbf{x}_k, \ldots;\mathbf{h}_m,
  \mathbf{y}_m,t)}_{\delta(\mathbf{v}-\mathbf{v}_k)f^{n\mathbf{u},m\mathbf{h}}(\mathbf{v}_1,\mathbf{x}_1;\ldots ;\mathbf{v}_k,\mathbf{x}_k; \ldots,\mathbf{v}_n,\mathbf{x}_n; \mathbf{b}_1,\mathbf{y}_1;\ldots \mathbf{b}_l,
  \mathbf{x}_k; \ldots;\mathbf{b}_m,\mathbf{y}_m,t)}\\
  &= \mathbf{v}_k \cdot \nabla_{\mathbf{x}_k} f^{n\mathbf{u},m\mathbf{h}}(\mathbf{v}_1,\mathbf{x}_1;\ldots ;\mathbf{v}_k,\mathbf{x}_k; \ldots,\mathbf{v}_n,\mathbf{x}_n; \mathbf{b}_1,\mathbf{y}_1;\ldots \mathbf{b}_l,\mathbf{x}_k; \ldots;
  \mathbf{b}_m,\mathbf{y}_m,t)\;,
\end{align}
where we made use of the so-called coincidence property of the PDF (\ref{eq:multi_pdf}). Further points can be fused by the same procedure. Moreover, bridging to velocity and magnetic field increment PDFs can be achieved by similar procedures to the ones described by Ulinich and Lyubimov for the case of hydrodynamic turbulence~\cite{Ulinich1969a}.
\section{Functional Formulation of MHD Turbulence}
In the previous section, we derived a hierarchy of evolution equations for PDFs in MHD turbulence which culminated in the generalized  evolution equation for the $(n+m)$-point PDF (\ref{eq:generalized}).
Owing to the fact that the MHD equations (\ref{eq:u}-\ref{eq:h}) form a classical field theory a complete description of the Eulerian statistics is also contained in the joint characteristic functional ~\cite{stanisic2012mathematical}
\begin{equation}
  \phi[\boldsymbol{\alpha}(\mathbf{x}),\boldsymbol{\beta}(\mathbf{x}),t]
  = \left \langle e^{i \int \textrm{d}\mathbf{x} \left [ \boldsymbol{\alpha}(\mathbf{x})\cdot \mathbf{u}(\mathbf{x},t) +\boldsymbol{\beta}(\mathbf{x})\cdot \mathbf{h}
  (\mathbf{x},t) \right ]}  \right \rangle\;.
  \label{eq:cumu_mhd}
\end{equation}
This functional can be used in order to project on both multi-point PDFs (\ref{eq:multi_pdf}) and multi-point correlations. E.g., the two-point cross helicity correlation function (\ref{eq:cuh}) can be obtained by two functional derivatives of the joint characteristic functional
\begin{equation}
  \left. \frac{\delta^2 \phi[\boldsymbol{\alpha}(\mathbf{x}),\boldsymbol{\beta}
  (\mathbf{x}),t]}{\delta (i \alpha_i(\mathbf{x}))\delta (i \beta_j(\mathbf{x'}))} \right|_{\boldsymbol{\beta}(\mathbf{x})=0}^{\boldsymbol{\alpha}(\mathbf{x})=0}=
  \left \langle u_i(\mathbf{x},t)
  h_j(\mathbf{x'},t) \right \rangle\;.
\end{equation}
In the following section, we briefly derive an evolution equation for the joint characteristic functional, whereas Sec.~\ref{sec:gauss} describes a Gaussian approximation for the joint characteristic functional (\ref{eq:cumu_mhd}).
\subsection{Evolution Equation for the Characteristic Functional in MHD Turbulence}
An evolution equation for the joint characteristic functional can be obtained by
deriving (\ref{eq:cumu_mhd}) with respect to time
\begin{align}
  \frac{\partial}{\partial t}
    \phi[\boldsymbol{\alpha}(\mathbf{x}),\boldsymbol{\beta}(\mathbf{x}),t]
    = i \int \textrm{d}\mathbf{x'} \left \langle
    \left [ \boldsymbol{\alpha}(\mathbf{x'})\cdot \frac{\partial \mathbf{u}(\mathbf{x'},t)}{\partial t} +\boldsymbol{\beta}(\mathbf{x'})\cdot \frac{\partial \mathbf{h}(\mathbf{x'},t)}{ \partial t} \right ]
    e^{i \int \textrm{d}\mathbf{x} \left [ \boldsymbol{\alpha}(\mathbf{x})\cdot \mathbf{u}(\mathbf{x},t) +\boldsymbol{\beta}(\mathbf{x})\cdot \mathbf{h}
    (\mathbf{x},t) \right ]} \right \rangle\;.
\end{align}
Inserting the MHD equations (\ref{eq:u}-\ref{eq:h}) yields
\begin{align}\label{eq:hopf1}
  \lefteqn{\frac{\partial}{\partial t}
    \phi[\boldsymbol{\alpha}(\mathbf{x}),\boldsymbol{\beta}(\mathbf{x}),t]}\\ \nonumber
    =& \int \textrm{d}\mathbf{x'} \alpha_j(\mathbf{x'})\left\{i \frac{\partial}{\partial x'_k}\left(\frac{\delta^2}{\delta \alpha_k(\mathbf{x'}) \delta \alpha_j(\mathbf{x'})}- \frac{\delta^2}{\delta \beta_k(\mathbf{x'}) \delta
    \beta_j(\mathbf{x'})} \right)+ \nu \frac{\partial^2}{\partial x'_k \partial x'_k}\frac{\delta}{\delta \alpha_j(\mathbf{x'})} \right.\\ \nonumber
    &+ \left. \frac{\partial}{\partial x'_j}
    \int
    \frac{\textrm{d}\mathbf{x''}}{4 \pi |\mathbf{x'}-\mathbf{x''}|}
    \frac{\partial^2}{\partial x''_k \partial x''_l} \left(\frac{\delta^2}{\delta \alpha_k(\mathbf{x''}) \delta \alpha_l(\mathbf{x''})}- \frac{\delta^2}{\delta \beta_k(\mathbf{x''}) \delta
    \beta_l(\mathbf{x''})} \right)    \right\}\phi[\boldsymbol{\alpha}(\mathbf{x}),\boldsymbol{\beta}(\mathbf{x}),t] \\ \nonumber
    &+\int \textrm{d}\mathbf{x'} \beta_j(\mathbf{x'})\left\{i
    \frac{\partial}{\partial x'_k}\left(\frac{\delta^2}{\delta \alpha_k(\mathbf{x'}) \delta \beta_j(\mathbf{x'})}- \frac{\delta^2}{\delta \beta_k(\mathbf{x'}) \delta
    \alpha_j(\mathbf{x'})} \right)+
    \lambda \frac{\partial^2}{\partial x'_k \partial x'_k}\frac{\delta}{\delta \beta_j(\mathbf{x'})} \right\}
    \phi[\boldsymbol{\alpha}(\mathbf{x}),\boldsymbol{\beta}(\mathbf{x}),t]\;.
\end{align}
Due to the incompressibility condition for the velocity field, the joint characteristic functional is invariant
under the transformation
\begin{equation}
\alpha_j(\mathbf{x}) =
  \tilde  \alpha_j(\mathbf{x}) +\frac{\partial}{\partial x_j} \psi(\mathbf{x})\;,
\end{equation}
and the pressure term in eq. (\ref{eq:hopf1})
can be eliminated, which yields
\begin{align}\label{eq:hopf2}
  \lefteqn{\frac{\partial}{\partial t}
    \phi[\boldsymbol{\alpha}(\mathbf{x}),\boldsymbol{\beta}(\mathbf{x}),t]}\\ \nonumber
    =& \int \textrm{d}\mathbf{x'} \tilde \alpha_j(\mathbf{x'})\left\{i \frac{\partial}{\partial x'_k}\left(\frac{\delta^2}{\delta \alpha_k(\mathbf{x'}) \delta \alpha_j(\mathbf{x'})}- \frac{\delta^2}{\delta \beta_k(\mathbf{x'}) \delta
    \beta_j(\mathbf{x'})} \right) + \nu \frac{\partial^2}{\partial x'_k\partial x'_k}\frac{\delta}{\delta \alpha_j(\mathbf{x'})} \right\}\phi[\boldsymbol{\alpha}(\mathbf{x}),\boldsymbol{\beta}(\mathbf{x}),t] \\ \nonumber
    &+\int \textrm{d}\mathbf{x'} \beta_j(\mathbf{x'})\left\{i
    \frac{\partial}{\partial x'_k}\left(\frac{\delta^2}{\delta \alpha_k(\mathbf{x'}) \delta \beta_j(\mathbf{x'})}- \frac{\delta^2}{\delta \beta_k(\mathbf{x'}) \delta
    \alpha_j(\mathbf{x'})} \right)+
    \lambda \frac{\partial^2}{\partial x'_k \partial x'_k}\frac{\delta}{\delta \beta_j(\mathbf{x'})} \right\}
    \phi[\boldsymbol{\alpha}(\mathbf{x}),\boldsymbol{\beta}(\mathbf{x}),t]\;.
\end{align}
This linear functional equation for the joint characteristic functional can be considered as the most compact statistical description of MHD turbulence
and is the analog of Hopf's functional formulation of hydrodynamic turbulence~\cite{stanisic2012mathematical,Grauer1996,Hopf1952}.
\subsection{Cumulant Expansion and the Implications of Vanishing Higher Order Cumulants}
\label{sec:gauss}
In the following, we consider the implications of the vanishing of cumulants higher than second order. Evidently, in the latter case, velocity and magnetic field are distributed according to a joint normal distribution. In fully developed MHD turbulence, however, such a scenario is quite unrealistic and empirical findings suggest strongly non-Gaussian behavior at small scales
which is oftentimes attributed to the occurrence of nearly singular current and vortex sheets~\cite{Friedrich2016,grauer1994scaling}. Nonetheless, as described in the next sections, we are solely interested in using the joint Gaussian approximation for mixed velocity and magnetic field conditional two-point moments in the single-point magnetic field evolution equation (\ref{eq:f_h}). Hence, as described in Appendix~\ref{app:gauss}, we approximate the joint
characteristic functional (\ref{eq:cumu_mhd}) according to
\begin{equation}
  \varphi[\boldsymbol{\alpha}(\mathbf{x}),\boldsymbol{\beta}(\mathbf{x}),t]
  = e^{-\frac{1}{2}\int \textrm{d}\mathbf{x'} \int \textrm{d}\mathbf{x''}  \left[\alpha_i(\mathbf{x'}) C^{\mathbf{u}\mathbf{u}}_{i\;j}(\mathbf{x'};\mathbf{x''},t) \alpha_j(\mathbf{x''})
  +2\alpha_i(\mathbf{x'},t)C^{\mathbf{u}\mathbf{h}}_{i\;j}(\mathbf{x'};\mathbf{x''},t) \beta_j(\mathbf{x''})
  +\beta_i(\mathbf{x'}) C^{\mathbf{h}\mathbf{h}}_{i\;j}
  (\mathbf{x'};\mathbf{x''},t) \beta_j(\mathbf{x''}) \right]}\;,
  \label{eq:cumu_gauss}
\end{equation}
where we introduced the two-point correlation functions
(\ref{eq:cuu}-\ref{eq:chh}). As a consequence, the entire multi-point statistics is determined by the correlation functions (\ref{eq:cuu}-\ref{eq:chh}).
\section{Closure of the Single-Point Magnetic Field PDF Equation and the Assumptions of Isotropy and Homogeneity.}
\label{sec:closure}
In this section, we rely on the characteristic functional (\ref{eq:cumu_gauss}) which was derived under the assumption of vanishing joint cumulants higher than order two. The closure method can thus be considered as a joint Gaussian closure assumption for the velocity and magnetic field.
Similar closures where discussed in the context of the multi-point vorticity statistics in two-dimensional turbulence~\cite{Friedrich2012}, as well as for the PDF of the velocity gradient tensor in three-dimensional turbulence~\cite{Wilczek2014}.
We reformulate the evolution equation of the single-point in terms of conditional expectation values
\begin{align}\nonumber
 \lefteqn{\frac{\partial}{\partial t}   f^{\mathbf{h}}(\mathbf{b}, \mathbf{x}, t)
 + \frac{\partial }{\partial x_i}  \left \langle u_i(\mathbf{x},t)|\mathbf{b}, \mathbf{x}, t \right\rangle
   f^{\mathbf{h}}(\mathbf{b}, \mathbf{x}, t)}\\
   \nonumber
   =& -\frac{\partial}{\partial b_i} \int \textrm{d}\mathbf{x'} \delta(\mathbf{x}-\mathbf{x'}) b_n \frac{\partial}{\partial x'_n} \left \langle u_i(\mathbf{x'},t)|\mathbf{b}, \mathbf{x}, t \right\rangle
   f^{\mathbf{h}}(\mathbf{b}, \mathbf{x}, t)\\
   & +\lambda \frac{\partial^2 }{\partial x_n^2} f^{\mathbf{h}}(\mathbf{b}, \mathbf{x}, t)-\lambda
   \frac{\partial^2 }{\partial b_i \partial b_j}
   \left \langle \frac{\partial h_i(\mathbf{x},t) }{\partial x_n} \frac{\partial h_j(\mathbf{x},t) }{\partial x_n}  \Bigg|\mathbf{b}, \mathbf{x}, t \right\rangle
   f^{\mathbf{h}}(\mathbf{b}, \mathbf{x}, t)\;,
   \label{eq:evo_h_cond}
\end{align}
where we imply summation over same indices. Moreover,
where we defined the conditional expectation values
\begin{align}\label{eq:u_b}
   \left \langle \mathbf{u}(\mathbf{x'},t)|\mathbf{b}, \mathbf{x}, t \right\rangle
   =&  \frac{\left \langle \mathbf{u}(\mathbf{x'},t) \delta(\mathbf{b}-\mathbf{h}(\mathbf{x},t)) \right\rangle}{\left \langle \delta(\mathbf{b}-\mathbf{h}( \mathbf{x},t)) \right \rangle}\;,\\
   \left \langle \frac{\partial h_i(\mathbf{x},t) }{\partial x_n} \frac{\partial h_j(\mathbf{x},t) }{\partial x_n}  \Bigg|\mathbf{b}, \mathbf{x}, t \right\rangle
   =&  \frac{\left \langle \frac{\partial h_i(\mathbf{x},t) }{\partial x_n} \frac{\partial h_j(\mathbf{x},t) }{\partial x_n} \delta(\mathbf{b}-\mathbf{h}(\mathbf{x},t)) \right\rangle}{\left \langle\delta(\mathbf{b}-\mathbf{h}(\mathbf{x},t))\right \rangle}\;,
   \label{eq:h_b}
\end{align}
and made use of the relation
\begin{equation}
 \frac{\partial^2 }{\partial x_n^2} f^{\mathbf{h}}(\mathbf{b}, \mathbf{x}, t)=
 - \frac{\partial }{\partial b_i}
 \left \langle \frac{\partial^2 h_i(\mathbf{x},t)}{\partial x_n^2}
 \delta(\mathbf{b}-\mathbf{h}(\mathbf{x},t))\right \rangle +
 \frac{\partial^2 }{\partial b_i \partial b_j}
 \left \langle \frac{\partial h_i(\mathbf{x},t)}{\partial x_n}
 \frac{\partial h_j(\mathbf{x},t)}{\partial x_n}
 \delta(\mathbf{b}-\mathbf{h}(\mathbf{x},t))\right \rangle\;.
\end{equation}
As shown in Appendix~\ref{app:gauss}, the first conditional expectation value (\ref{eq:u_b}) can be calculated from the joint Gaussian characteristic functional according to
\begin{equation}
  \label{eq:cond_u}
  \left \langle u_i(\mathbf{x'},t)|\mathbf{b}, \mathbf{x}, t \right\rangle
  =   C^{\mathbf{u}\mathbf{h}}_{i\;j}(\mathbf{x'};\mathbf{x},t) C^{\mathbf{h}\mathbf{h}}_{j\;k}(\mathbf{x};\mathbf{x},t)^{-1} b_k\;
\end{equation}
where $C^{\mathbf{h}\mathbf{h}}_{j\;k}(\mathbf{x};\mathbf{x'},t)^{-1}$ denotes the inverse of the two-point tensor (\ref{eq:chh}). Furthermore, under the assumption of translational invariance of the fields, we can neglect derivatives of single-point quantities  $\nabla_{\mathbf{x}} \cdot \langle \ldots \rangle$, such as the second term on the l.h.s. of eq. (\ref{eq:evo_h_cond}).
In addition, two-point correlations will only depend on the separation vector $\mathbf{r}=\mathbf{x}-\mathbf{x'}$. Next to homogeneity, we also want to impose isotropy, i.e., rotational invariance. Here, special attention has to be paid to the fact that unlike $\mathbf{u}$, $\mathbf{h}$ is an axial vector. Therefore, correlations which contain an odd number of magnetic  field components such as (\ref{eq:cond_u})
are skew tensors~\cite{Chandrasekhar1950,Robertson1940,chandra:1951}. Hence, the correlation tensors that enter the conditional expectation values in eq. (\ref{eq:cond_u}) read
\begin{align}
  \label{eq:cij}
  C^{\mathbf{u}\mathbf{h}}_{i\;j}(\mathbf{r},t) &=   C^{\mathbf{u}\mathbf{h}}(r,t)\varepsilon_{ijk}r_k \\
  C^{\mathbf{h}\mathbf{h}}_{i\;j}(0,t) &= h_{rms}^2 \delta_{ij}\;, 
\end{align}
where $h_{rms}$ denotes the root mean square magnetic field $\sqrt{\langle h^2 \rangle}$.
Furthermore, the assumption of isotropy implies that the single-point PDF solely depends on the absolute of the magnetic field $b=|\mathbf{b}|$. This PDF is related to the PDF of the absolute value of the magnetic field $f^{h}(b,t)$
according to~\cite{Wilczek2011c}
\begin{equation}
   f^{h}(b,t)= 4 \pi b^2 f^{\mathbf{h}}(\mathbf{b}, \mathbf{x}, t)\;.
\end{equation}
Therefore, the remaining nonlinear term, i.e., the first term on the r.h.s. of eq. (\ref{eq:evo_h_cond})
yields
\begin{align}\nonumber
  \lefteqn{-\nabla_{\mathbf{b}} \cdot  \int \textrm{d}\mathbf{r} \delta(\mathbf{r}) \left(\mathbf{b}\cdot \nabla_\mathbf{r} \right) \left \langle \mathbf{u}(\mathbf{x}+\mathbf{r},t)|\mathbf{b}, \mathbf{x}, t
   \right\rangle
  f^{\mathbf{h}}(b, \mathbf{x}, t)}\\ \nonumber
 =&- h_{rms}^{-2} \frac{\partial}{\partial b_i} \int \textrm{d}\mathbf{r} \delta(\mathbf{r})b_n \frac{\partial}{\partial r_n} C^{\mathbf{u}\mathbf{h}}(r,t)\varepsilon_{ijl}r_l  \delta_{jk} b_k
f^{\mathbf{h}}(b, \mathbf{x}, t)\\ \nonumber
 =& - h_{rms}^{-2} \int \textrm{d}\mathbf{r} \delta(\mathbf{r}) \left(\frac{\partial C^{\mathbf{u}\mathbf{h}}(r,t)}{\partial r}\frac{r_n}{r} \varepsilon_{ijl}r_l +C^{\mathbf{u}\mathbf{h}}(r,t) \varepsilon_{ijn}
  \right)\\
 &\times\left(\delta_{in}b_jf^{\mathbf{h}}(b, \mathbf{x}, t)+ \delta_{ij}b_n f^{\mathbf{h}}(b, \mathbf{x}, t) +b_j b_n \frac{b_i}{b} \frac{\partial f^{\mathbf{h}}(\mathbf{b}, \mathbf{x}, t)}{\partial b} \right)
 =0\;.
 \label{eq:non}
\end{align}
where we made use of the general relation $\frac{\partial}{\partial r_n}g(r)= \frac{\partial g(r)}{\partial r}\frac{r_n}{r}$
and used the properties of the Levi-Civita tensor, e.g.,
$\varepsilon_{ijl}r_l r_i=0$, and $\varepsilon_{iil}=0$.
Remarkably, the joint Gaussian approximation combined with the assumptions of homogeneity and isotropy leads to vanishing nonlinear transfer terms in the single-point PDF equation. In other words, growth of magnetic field fluctuations - in a setting with translational invariance and no preferred direction - requires non-vanishing
higher-order cumulants, e.g., generated by non-vanishing third order moments.

The remaining term involves the conditional expectation value (\ref{eq:h_b}) which simplifies to
\begin{equation}
  \left \langle \frac{\partial h_i(\mathbf{x},t) }{\partial x_n} \frac{\partial h_j(\mathbf{x},t) }{\partial x_n}  \Bigg|\mathbf{b}, \mathbf{x}, t \right\rangle=-
  \frac{\left \langle \varepsilon^{\mathbf{h}\mathbf{h}} \right \rangle}{3\lambda} \delta_{ij}\;,
\end{equation}
where we have introduced the average of the local magnetic energy dissipation rate according to
\begin{equation}
  \left \langle \varepsilon^{\mathbf{h}\mathbf{h}} \right \rangle=\frac{\lambda}{2}
  \left \langle \left[\frac{\partial h_i(\mathbf{x},t) }{\partial x_j}+ \frac{\partial h_j(\mathbf{x},t) }{\partial x_i}\right]^2 \right \rangle \;.
\end{equation}
Consequently, eq. (\ref{eq:evo_h_cond}) reduces to
\begin{equation}
  \frac{\partial}{\partial t} \tilde f^{h}(b,t)
  = -\frac{\partial}{\partial b} \frac{ 2 \left \langle \varepsilon^{\mathbf{h}\mathbf{h}} \right \rangle}{3b}\tilde f^{h}(b,t) + \frac{\partial^2}{\partial b^2}  \frac{\left \langle \varepsilon^{\mathbf{h}\mathbf{h}} \right \rangle}{3}
  \tilde f^{h}(b,t)\;,
  \label{eq:norm}
\end{equation}
Accordingly, taking the second moment of this equation yields the temporal evolution for the magnetic energy
\begin{equation}
  \frac{\partial }{\partial t} E_{mag}(t)
  = \frac{1}{2}\int \textrm{d}b b^2  \frac{\partial }{\partial t}f^{h}(b,t)
  = -  \left \langle \varepsilon^{\mathbf{h}\mathbf{h}} \right \rangle\;.
\end{equation}
Due to the vanishing of the nonlinear term (\ref{eq:non}), eq. (\ref{eq:norm}) possesses no transfer term, which again is a direct consequence of the isotropy/homogeneity assumption and the neglect of higher order cumulants in the joint characteristic functional (\ref{eq:cumu_mhd}). A possible way of restoring this missing term within the same joint Gaussian approximation consist in allowing for a mean magnetic field $\mathbf{H}(t)=\langle \mathbf{h}(\mathbf{x},t) \rangle$, which fluctuates about the mean direction $\boldsymbol \lambda =\langle \mathbf{H}(t)/H(t) \rangle_t$. In this case,
we necessarily have to deal with axisymmetric MHD turbulence and
the skew symmetric tensor (\ref{eq:cij}) gets modified to~\cite{Chandrasekhar1950}
\begin{equation}
 C^{\mathbf{u}\mathbf{h}}_{i\;j}(\mathbf{r},t) =   C^{(1)}(r,t)\varepsilon_{ijk}r_k+
 C^{(2)}(r,t)\lambda_j \varepsilon_{ilm}\lambda_l r_m
 +C^{(3)}(r,t)r_j \varepsilon_{ilm}\lambda_l r_m\;,
\end{equation}
which allows for a non-vanishing transfer term in eq. (\ref{eq:evo_h_cond}). The implications of this theory for dynamo theory  and will be discussed in a forthcoming publication.
\section{Discussion}
We have presented a comprehensive statistical description of MHD turbulence. First, we formulated   a hierarchy of evolution equations for multi-point probability density functions that can be considered as the analogon to the multi-point PDF hierarchy derived by Lundgren~\cite{Lundgren1967}. In comparison to the latter hierarchy of hydrodynamic turbulence, where unclosed terms are due to pressure and viscous contributions, the Alfv\'en effect introduces additional unclosed terms. Furthermore, we discussed the relation of the PDF hierarchy to the moment hierarchy of Chandrasekhar~\cite{chandra:1951} and presented a compact formulation on the basis of a joint characteristic functional in MHD turbulence~\cite{stanisic2012mathematical}. A joint Gaussian approximation for this characteristic functional has been used in order to close two-point terms in the evolution equation for the single-point
magnetic field PDF.

It will be a task for the future to directly assess these conditional averages by direct numerical simulations, similarly to the case of hydrodynamic turbulence~\cite{Friedrich2012,Wilczek2014} and further quantify deviations from Gaussianity and their origin in the dynamics of MHD turbulence.



\funding{I acknowledge funding from the Humboldt Foundation within a Feodor-Lynen fellowship. Moreover, I benefited from financial support through the Project IDEXLYON of the University of Lyon in the framework of the French program ``Programme Investissements d'Avenir'' (ANR-16-IDEX-0005).}
\acknowledgments{I want to thank Alain Pumir for interesting discussions.}

\conflictsofinterest{The author declares no conflict of interest.}

\appendixtitles{no} 
\appendix
\section{Derivation of the PDF hierarchy}
\label{app:deriv_hierarchy}
This appendix discusses the derivation of the evolution equations for the single-point velocity  (\ref{eq:evo_u}) and magnetic field PDF (\ref{eq:evo_h}).
\subsection{Terms of the evolution equation for the single-point velocity PDF}
\label{app:f_u}
The first nonlinear term in eq. (\ref{eq:u}) can be related to the single-point PDF by the following manipulation
\begin{align}\nonumber
  \lefteqn{\nabla_{\mathbf{v}} \cdot \left \langle  \left[\mathbf{u}(\mathbf{x},t) \cdot \nabla_{\mathbf{x}}\mathbf{u}(\mathbf{x},t)\right]\delta(\mathbf{v}-\mathbf{u}(\mathbf{x},t)) \right \rangle}\\
   &=-
  \left \langle  \mathbf{u}(\mathbf{x},t) \cdot \nabla_{\mathbf{x}}\delta(\mathbf{v}-\mathbf{u}(\mathbf{x},t)) \right \rangle
  = -\nabla_{\mathbf{x}} \cdot \left \langle  \mathbf{u}(\mathbf{x},t)\delta(\mathbf{v}-\mathbf{u}(\mathbf{x},t)) \right \rangle = - \mathbf{v} \cdot \nabla_{\mathbf{x}}
  \underbrace{\left \langle  \delta(\mathbf{v}-\mathbf{u}(\mathbf{x},t)) \right \rangle}_{=  f^{\mathbf{u}}(\mathbf{v}, \mathbf{x}, t)}\;,
  \label{eq:evo_u1}
\end{align}
where we made use of the incompressibility of the velocity field after the second equality and used the so-called sifting property of the $\delta$-distribution, $\mathbf{u}(\mathbf{x},t) \delta(\mathbf{v}-\mathbf{u}(\mathbf{x},t)) =\mathbf{v} \delta(\mathbf{v}-\mathbf{u}(\mathbf{x},t))$, after the last equality.
The second nonlinear term, however, involves the magnetic field and has to be related to a two-point quantity
\begin{align}\nonumber
  \lefteqn{-\nabla_{\mathbf{v}} \cdot \left \langle  \left[\mathbf{h}(\mathbf{x},t) \cdot \nabla_{\mathbf{x}}\mathbf{h}(\mathbf{x},t) \right]\delta(\mathbf{v}-\mathbf{u}(\mathbf{x},t)) \right \rangle}\\ \nonumber
   &=- \frac{\partial}{\partial v_i} \left \langle \left[ h_n(\mathbf{x},t)\frac{\partial}{\partial x_n} h_i(\mathbf{x},t)\right]\delta(\mathbf{v}-\mathbf{u}(\mathbf{x},t)) \right \rangle\\
   \nonumber
   &= - \frac{\partial}{\partial v_i} \int \textrm{d}\mathbf{x'}\delta(\mathbf{x}-\mathbf{x'})\frac{\partial}{\partial x'_n} \left \langle  h_n(\mathbf{x'},t) h_i(\mathbf{x'},t)\delta(\mathbf{v}-\mathbf{u}(\mathbf{x},t)) \right \rangle\\ \nonumber
   &= -\frac{\partial}{\partial v_i} \int \textrm{d}\mathbf{x'}\delta(\mathbf{x}-\mathbf{x'}) \frac{\partial}{\partial x'_n} \int \textrm{d}\mathbf{b}\; b_n b_i
   \underbrace{\left \langle   \delta(\mathbf{v}-\mathbf{u}(\mathbf{x},t)) \delta(\mathbf{b}-\mathbf{h}(\mathbf{x'},t))  \right \rangle}_{
    f^{\mathbf{u}\mathbf{h}}(\mathbf{v}, \mathbf{x}; \mathbf{b}, \mathbf{x'}, t)
   }\\
  &= -\nabla_{\mathbf{v}} \cdot  \int \textrm{d}\mathbf{x'}\delta(\mathbf{x}-\mathbf{x'}) \int \textrm{d}\mathbf{b}\; \mathbf{b}\cdot \nabla_{\mathbf{x'}}
  \mathbf{b}      f^{\mathbf{u}\mathbf{h}}(\mathbf{v}, \mathbf{x}; \mathbf{b}, \mathbf{x'}, t)\;,
  \label{eq:evo_u2}
\end{align}
where we implied summation over equal indices.
The next term involves nonlocal pressure contributions and therefore, again, two-point quantities. The total pressure field can be obtained by taking the divergence of eq. (\ref{eq:u}) which yields the Poisson equation
\begin{equation}
  - \nabla_{\mathbf x}^2 p(\mathbf{x},t)=
   \nabla_{\mathbf x} \cdot [\mathbf{u}(\mathbf{x},t) \cdot \nabla_{\mathbf{x}}\mathbf{u}(\mathbf{x},t)
   - \mathbf{h}(\mathbf{x},t) \cdot \nabla_{\mathbf{x}}\mathbf{h}(\mathbf{x},t)]\;,
\end{equation}
which can be solved by the usual Green function method according to
\begin{equation}
  p(\mathbf{x},t) = \frac{1}{4\pi} \int \textrm{d}\mathbf{x'}\frac{1}{|\mathbf{x}-\mathbf{x'}|}
  \nabla_{\mathbf x'} \cdot [\mathbf{u}(\mathbf{x'},t) \cdot \nabla_{\mathbf{x'}}\mathbf{u}(\mathbf{x'},t)
  - \mathbf{h}(\mathbf{x'},t) \cdot \nabla_{\mathbf{x'}}\mathbf{h}(\mathbf{x'},t)]
    \;,
\end{equation}
where we assumed an infinite domain without further boundary conditions.
Accordingly, the pressure term that enters eq. (\ref{eq:evo_u}) reads
\begin{align} \nonumber
\lefteqn{\nabla_{\mathbf{v}} \cdot \left \langle \left[ \nabla_{\mathbf{x}}p(\mathbf{x},t) \right]\delta(\mathbf{v}-\mathbf{u}(\mathbf{x},t)) \right \rangle}\\ \nonumber
&= \frac{1}{4\pi} \nabla_{\mathbf{v}}\cdot
\int \textrm{d}\mathbf{x'} \left[\nabla_{\mathbf{x}} \frac{1}{|\mathbf{x}-\mathbf{x'}|}\right]  \frac{\partial^2}{\partial x'_i \partial x'_n}
\left \langle \left[u_n(\mathbf{x'},t)u_i(\mathbf{x'},t)- h_n(\mathbf{x'},t) h_i(\mathbf{x'},t) \right]\delta(\mathbf{v}-\mathbf{u}(\mathbf{x},t)) \right\rangle\\ \nonumber
&= \frac{1}{4\pi} \nabla_{\mathbf{v}}\cdot
\int \textrm{d}\mathbf{x'} \left[\nabla_{\mathbf{x}} \frac{1}{|\mathbf{x}-\mathbf{x'}|}\right]  \frac{\partial^2}{\partial x'_i \partial x'_n}
\int \textrm{d}\mathbf{v'} v'_n v'_i \left \langle \delta(\mathbf{v}-\mathbf{u}(\mathbf{x},t))
\delta(\mathbf{v'}-\mathbf{u}(\mathbf{x'},t)) \right\rangle \\ \nonumber
&~~-\frac{1}{4\pi} \nabla_{\mathbf{v}}\cdot
\int \textrm{d}\mathbf{x'} \left[\nabla_{\mathbf{x}} \frac{1}{|\mathbf{x}-\mathbf{x'}|}\right]  \frac{\partial^2}{\partial x'_i \partial x'_n}
\int \textrm{d}\mathbf{b'} b'_n b'_i \left \langle
\delta(\mathbf{v}-\mathbf{u}(\mathbf{x},t))
\delta(\mathbf{b'}-\mathbf{h}(\mathbf{x'},t)) \right\rangle\\ \nonumber
&= \frac{1}{4\pi} \nabla_{\mathbf{v}}\cdot
\int \textrm{d}\mathbf{x'} \left[\nabla_{\mathbf{x}} \frac{1}{|\mathbf{x}-\mathbf{x'}|}\right] \int \textrm{d}\mathbf{v'} \left(\mathbf{v'}\cdot \nabla_{\mathbf{x'}} \right)^2
f^{\mathbf{u}\mathbf{u}}(\mathbf{v}, \mathbf{x}; \mathbf{v'}, \mathbf{x'}, t)\\
&~~ -\frac{1}{4\pi} \nabla_{\mathbf{v}}\cdot
\int \textrm{d}\mathbf{x'} \left[\nabla_{\mathbf{x}} \frac{1}{|\mathbf{x}-\mathbf{x'}|}\right] \int \textrm{d}\mathbf{b} \left(\mathbf{b}\cdot \nabla_{\mathbf{x'}} \right)^2
f^{\mathbf{u}\mathbf{h}}(\mathbf{v}, \mathbf{x}; \mathbf{b}, \mathbf{x'}, t)\;.
\label{eq:evo_u3}
\end{align}
Finally, the viscous term reads
\begin{align}\nonumber
  \lefteqn{-\nu \nabla_{\mathbf{v}} \cdot \left \langle \left[ \nabla_{\mathbf{x}}^2 \mathbf{u} (\mathbf{x},t) \right]\delta(\mathbf{v}-\mathbf{u}(\mathbf{x},t)) \right \rangle}\\ \nonumber
  &=
-\nu \nabla_{\mathbf{v}} \cdot \int \textrm{d}\mathbf{x'} \delta(\mathbf{x}-\mathbf{x'}) \nabla_{\mathbf{x'}}^2 \left \langle \mathbf{u} (\mathbf{x'},t))
\delta(\mathbf{v}-\mathbf{u}(\mathbf{x},t)) \right \rangle \\
&=
-\nu \nabla_{\mathbf{v}} \cdot \int \textrm{d}\mathbf{x'} \delta(\mathbf{x}-\mathbf{x'}) \nabla_{\mathbf{x'}}^2 \int \textrm{d}\mathbf{v'} \mathbf{v'}
\underbrace{\left \langle
\delta(\mathbf{v}-\mathbf{u}(\mathbf{x},t)) \delta(\mathbf{v'}-\mathbf{u}(\mathbf{x'},t)) \right\rangle}_{f^{\mathbf{u}\mathbf{u}}(\mathbf{v}, \mathbf{x}; \mathbf{v'}, \mathbf{x'}, t)}
\;,
\label{eq:evo_u4}
\end{align}
Inserting the terms (\ref{eq:evo_u1}), (\ref{eq:evo_u2}), (\ref{eq:evo_u3}), and (\ref{eq:evo_u4}) into eq. (\ref{eq:evo_u_init}) yields the evolution equation for the single-point velocity PDF (\ref{eq:evo_u}).
\subsection{Terms of the evolution equation for the single-point velocity PDF}
\label{app:f_h}
The first nonlinear term from the induction equation (\ref{eq:h}) can be related to a single-point mixed quantity
\begin{align}\nonumber
  \lefteqn{\nabla_{\mathbf{b}} \cdot \left \langle  \left[\mathbf{u}(\mathbf{x},t) \cdot \nabla_{\mathbf{x}}\mathbf{h}(\mathbf{x},t)\right]\delta(\mathbf{b}-\mathbf{h}(\mathbf{x},t)) \right \rangle}\\ \nonumber
   &=-
  \left \langle  \mathbf{u}(\mathbf{x},t) \cdot \nabla_{\mathbf{x}}\delta(\mathbf{b}-\mathbf{h}(\mathbf{x},t)) \right \rangle
  = -\nabla_{\mathbf{x}} \cdot \left \langle  \mathbf{u}(\mathbf{x},t) \delta(\mathbf{b}-\mathbf{h}(\mathbf{x},t)) \right \rangle \\
  &= -\nabla_{\mathbf{x}} \cdot \int \textrm{d} \mathbf{v} \mathbf{v}
  \left \langle \delta(\mathbf{v}- \mathbf{u}(\mathbf{x},t)) \delta(\mathbf{b}-\mathbf{h}(\mathbf{x},t)) \right \rangle =
  \int \textrm{d}\mathbf{v} \mathbf{v}\cdot \nabla_{\mathbf{x}}
    f^{\mathbf{u}\mathbf{h}}(\mathbf{v}, \mathbf{x};\mathbf{b}, \mathbf{x}, t)\;.
    \label{eq:evo_h1}
\end{align}
The second term, however, has to be related to the two-point PDF (\ref{eq:f_uh}) according to
\begin{align}\nonumber
  \lefteqn{-\nabla_{\mathbf{b}} \cdot \left \langle  \left[\mathbf{h}(\mathbf{x},t) \cdot \nabla_{\mathbf{x}}\mathbf{u}(\mathbf{x},t)\right]\delta(\mathbf{b}-\mathbf{h}(\mathbf{x},t)) \right \rangle}\\ \nonumber
    &=- \frac{\partial}{\partial b_i} \left \langle \left[ h_n(\mathbf{x},t)\frac{\partial}{\partial x_n} u_i(\mathbf{x},t)\right]\delta(\mathbf{b}-\mathbf{h}(\mathbf{x},t)) \right \rangle\\
    \nonumber
    &= - \frac{\partial}{\partial b_i} \int \textrm{d}\mathbf{x'}\delta(\mathbf{x}-\mathbf{x'})\frac{\partial}{\partial x'_n} \left \langle  h_n(\mathbf{x},t) u_i(\mathbf{x'},t)\delta(\mathbf{b}-\mathbf{h}(\mathbf{x},t)) \right \rangle\\ \nonumber
    &= -\frac{\partial}{\partial b_i} \int \textrm{d}\mathbf{x'}\delta(\mathbf{x}-\mathbf{x'}) \frac{\partial}{\partial x'_n} \int \textrm{d}\mathbf{v}\; b_n v_i
    \left \langle   \delta(\mathbf{v}-\mathbf{u}(\mathbf{x'},t)) \delta(\mathbf{b}-\mathbf{h}(\mathbf{x},t))  \right \rangle\\
    &= -\nabla_{\mathbf{b}} \cdot \int \textrm{d}\mathbf{x'}\delta(\mathbf{x}-\mathbf{x'})  \int \textrm{d}\mathbf{v}\left(\mathbf{b}\cdot \nabla_\mathbf{x'} \right)\mathbf{v}
    f^{\mathbf{u}\mathbf{h}}(\mathbf{v}, \mathbf{x'};\mathbf{b}, \mathbf{x}, t)
\end{align}
Finally, the viscous term that enters the evolution equation for the one-point magnetic field PDF can be treated in the same way as eq. (\ref{eq:evo_u4}).
\section{Relation to the Moment Hierarchy Derived by Chandrasekhar}
\label{app:chandra}
In this section we want to derive the evolution equation for the moments (\ref{eq:cuu}-\ref{eq:chh}) directly from the PDF equations.
\subsection{Evolution Equation for the Two-Point Velocity Field Correlation Tensor}
The evolution equation for the two-point velocity PDF (\ref{eq:f_uu}) can be obtained from the generalized equation (\ref{eq:generalized}) according to
\begin{align}\label{eq:evo_uu}
\lefteqn{ \left[ \frac{\partial}{\partial t} + \mathbf{v} \cdot \nabla_{\mathbf{x}} + \mathbf{v'} \cdot \nabla_{\mathbf{x'}}\right]
f^{\mathbf{u}\mathbf{u}}(\mathbf{v}, \mathbf{x};
\mathbf{v'}, \mathbf{x'}, t) }\\ \nonumber
= &-\left[\nabla_{\mathbf{v}} \cdot  \int \textrm{d}\mathbf{x''}\delta(\mathbf{x''}-\mathbf{x})+
\nabla_{\mathbf{v'}} \cdot  \int \textrm{d}\mathbf{x''}\delta(\mathbf{x''}-\mathbf{x'}) \right] \int \textrm{d}\mathbf{b} \left(\mathbf{b}\cdot \nabla_{\mathbf{x''}}\right)
\mathbf{b}      f^{\mathbf{u}\mathbf{u}\mathbf{h}}(\mathbf{v}, \mathbf{x};
\mathbf{v'}, \mathbf{x'}; \mathbf{b}, \mathbf{x''}, t)\\ \nonumber
&-\nu \left[\nabla_{\mathbf{v}} \cdot  \int \textrm{d}\mathbf{x''}\delta(\mathbf{x''}-\mathbf{x})+
\nabla_{\mathbf{v'}} \cdot  \int \textrm{d}\mathbf{x''}\delta(\mathbf{x''}-\mathbf{x'}) \right]  \nabla_{\mathbf{x''}}^2 \int \textrm{d}\mathbf{v''} \mathbf{v''}
f^{\mathbf{u} \mathbf{u}\mathbf{u}}(\mathbf{v''}, \mathbf{x''}; \mathbf{v}, \mathbf{x};\mathbf{v'}, \mathbf{x'}, t)\;,
\end{align}
where we already neglected pressure contributions due to the fact that they vanish on the basis of homogeneity at this stage of the moment hierarchy discussed in the following (nonetheless, they do not vanish for higher order correlations~\cite{Friedrich2016}). Taking the moments
\begin{equation}
  \langle v_i(\mathbf{x},t)v_j(\mathbf{x'},t) \rangle= \int \textrm{d}\mathbf{v} v_{i}
  \int \textrm{d}\mathbf{v'} v'_{j}
  f^{\mathbf{u}\mathbf{u}}(\mathbf{v}, \mathbf{x};
  \mathbf{v'}, \mathbf{x'}, t)\;,
\end{equation}
of eq. (\ref{eq:evo_uu}) yields
\begin{align}
  \lefteqn{\frac{\partial}{\partial t}
  \langle u_i(\mathbf{x},t)u_j(\mathbf{x'},t)
  \rangle
  + \frac{\partial}{\partial x_{n}}
    \langle u_n(\mathbf{x},t) u_i(\mathbf{x},t)u_j(\mathbf{x'},t)\rangle
  + \frac{\partial}{\partial x'_{n}}
    \langle u_n(\mathbf{x'},t) u_i(\mathbf{x},t)u_j(\mathbf{x'},t)\rangle} \\ \nonumber
    &= \frac{\partial}{\partial x_{n}}
      \langle h_n(\mathbf{x},t) h_i(\mathbf{x},t)u_j(\mathbf{x'},t) \rangle
    + \frac{\partial}{\partial x'_{n}}
      \langle h_n(\mathbf{x'},t) u_i(\mathbf{x},t)h_j(\mathbf{x'},t)\rangle
      + \nu [\nabla_{\mathbf{x}}^2+\nabla_{\mathbf{x'}}^2]
      \langle u_i(\mathbf{x},t)u_j(\mathbf{x'},t)
      \rangle\;.
\end{align}
where we performed partial integration with respect to $\mathbf{v}$ and $\mathbf{v'}$ on the r.h.s. of eq. (\ref{eq:evo_uu}).
\subsection{Evolution Equation for the Two-Point Cross Helicity Correlation Tensor}
The evolution equation for the two-point PDF (\ref{eq:f_uh}) reads
\begin{align}\nonumber
  \lefteqn{\left[\frac{\partial}{\partial t} +\mathbf{v}\cdot \nabla_{\mathbf{x}}\right]     f^{\mathbf{u}\mathbf{h}}(\mathbf{v}, \mathbf{x};\mathbf{b}, \mathbf{x'}, t)
  +\int \textrm{d}\mathbf{v'} \mathbf{v'}\cdot \nabla_{\mathbf{x'}}
    f^{\mathbf{u}\mathbf{u}\mathbf{h}}(\mathbf{v'}, \mathbf{x'}; \mathbf{v}, \mathbf{x};\mathbf{b}, \mathbf{x'}, t)}\\
    \nonumber
    =& -\nabla_{\mathbf{v}} \cdot \int \textrm{d}\mathbf{x''}\delta(\mathbf{x}-\mathbf{x''})  \int \textrm{d}\mathbf{b'} \left(\mathbf{b'}\cdot \nabla_\mathbf{x''} \right) \mathbf{b'}
    f^{\mathbf{u}\mathbf{h}\mathbf{h}}(\mathbf{v}, \mathbf{x};\mathbf{b'}, \mathbf{x''};\mathbf{b}, \mathbf{x'}, t)\\ \nonumber
    &-\nabla_{\mathbf{b}} \cdot \int \textrm{d}\mathbf{x''}\delta(\mathbf{x'}-\mathbf{x''})  \int \textrm{d}\mathbf{v'} \left(\mathbf{b}\cdot \nabla_\mathbf{x''} \right) \mathbf{v'}
    f^{\mathbf{u}\mathbf{u}\mathbf{h}}(\mathbf{v'}, \mathbf{x''};\mathbf{v}, \mathbf{x};\mathbf{b}, \mathbf{x'}, t)\\ \nonumber
    &-\nu \nabla_{\mathbf{v}} \cdot \int \textrm{d}\mathbf{x''} \delta(\mathbf{x}-\mathbf{x''}) \nabla_{\mathbf{x''}}^2 \int \textrm{d}\mathbf{v'} \mathbf{v'}
    f^{\mathbf{u}\mathbf{u}\mathbf{h}}(\mathbf{v'}, \mathbf{x''};\mathbf{v}, \mathbf{x};\mathbf{b}, \mathbf{x'}, t)\\
    & -\lambda \nabla_{\mathbf{b}} \cdot \int \textrm{d}\mathbf{x''} \delta(\mathbf{x'}-\mathbf{x''}) \nabla_{\mathbf{x''}}^2 \int \textrm{d}\mathbf{b'} \mathbf{b'}
    f^{\mathbf{u}\mathbf{h}\mathbf{h}}(\mathbf{v}, \mathbf{x};\mathbf{b'}, \mathbf{x''};\mathbf{b}, \mathbf{x'}, t)\;,
    \label{eq:evo_uh}
\end{align}
where we again dropped pressure contributions.
Taking the moments
\begin{equation}
  \langle u_i(\mathbf{x},t)h_j(\mathbf{x'},t) \rangle= \int \textrm{d}\mathbf{v} v_{i}
  \int \textrm{d}\mathbf{b} b_{j}
  f^{\mathbf{u}\mathbf{h}}(\mathbf{v}, \mathbf{x};
  \mathbf{b}, \mathbf{x'}, t)\;,
\end{equation}
of eq. (\ref{eq:evo_uh}) yields
\begin{align}
  \lefteqn{\frac{\partial}{\partial t}
  \langle u_i(\mathbf{x},t)h_j(\mathbf{x'},t)
  \rangle
  + \frac{\partial}{\partial x_{n}}
    \langle u_n(\mathbf{x},t) u_i(\mathbf{x},t)h_j(\mathbf{x'},t)\rangle
  + \frac{\partial}{\partial x'_{n}}
    \langle u_n(\mathbf{x'},t) u_i(\mathbf{x},t)h_j(\mathbf{x'},t)\rangle} \\ \nonumber
    &= \frac{\partial}{\partial x_{n}}
      \langle h_n(\mathbf{x},t) h_i(\mathbf{x},t)h_j(\mathbf{x'},t) \rangle
    + \frac{\partial}{\partial x'_{n}}
      \langle h_n(\mathbf{x'},t) u_i(\mathbf{x},t)u_j(\mathbf{x'},t)\rangle
      +  [\nu\nabla_{\mathbf{x}}^2+\lambda\nabla_{\mathbf{x'}}^2]
      \langle u_i(\mathbf{x},t)h_j(\mathbf{x'},t)
      \rangle\;.
\end{align}
\subsection{Evolution Equation for the Two-Point Magnetic Field Correlation Tensor}
Finally, the evolution equation for the two-point magnetic field PDF (\ref{eq:f_hh}) reads
\begin{align}\nonumber
 \lefteqn{\frac{\partial}{\partial t}   f^{\mathbf{h}\mathbf{h}}(\mathbf{b}, \mathbf{x};\mathbf{b'}, \mathbf{x'}, t)
 +\int \textrm{d}\mathbf{v} \mathbf{v}\cdot \nabla_{\mathbf{x}}
   f^{\mathbf{u}\mathbf{h}\mathbf{h}}(\mathbf{v}, \mathbf{x};\mathbf{b}, \mathbf{x};\mathbf{b'}, \mathbf{x'}, t)+\int \textrm{d}\mathbf{v} \mathbf{v}\cdot \nabla_{\mathbf{x'}}
     f^{\mathbf{u}\mathbf{h}\mathbf{h}}(\mathbf{v}, \mathbf{x'};\mathbf{b}, \mathbf{x},\mathbf{b'}, \mathbf{x'}, t)}\\ \nonumber
   =& -\nabla_{\mathbf{b}} \cdot \int \textrm{d}\mathbf{x''}\delta(\mathbf{x}-\mathbf{x''})  \int \textrm{d}\mathbf{v} \left(\mathbf{b}\cdot \nabla_\mathbf{x''} \right) \mathbf{v}
   f^{\mathbf{u}\mathbf{h}\mathbf{h}}(\mathbf{v}, \mathbf{x''};\mathbf{b}, \mathbf{x}; \mathbf{b'}, \mathbf{x'}, t)\\ \nonumber
   &-\nabla_{\mathbf{b'}} \cdot \int \textrm{d}\mathbf{x''}\delta(\mathbf{x'}-\mathbf{x''})  \int \textrm{d}\mathbf{v} \left(\mathbf{b'}\cdot \nabla_\mathbf{x''} \right) \mathbf{v}
   f^{\mathbf{u}\mathbf{h}\mathbf{h}}(\mathbf{v}, \mathbf{x''};\mathbf{b}, \mathbf{x}; \mathbf{b'}, \mathbf{x'}, t)\\
   &-\lambda \left[\nabla_{\mathbf{b}} \cdot  \int \textrm{d}\mathbf{x}\delta(\mathbf{x''}-\mathbf{x})+
   \nabla_{\mathbf{b'}} \cdot  \int \textrm{d}\mathbf{x}\delta(\mathbf{x''}-\mathbf{x'}) \right]  \nabla_{\mathbf{x''}}^2 \int \textrm{d}\mathbf{b''} \mathbf{b''}
   f^{\mathbf{h} \mathbf{h}\mathbf{h}}(\mathbf{b''}, \mathbf{x''}; \mathbf{b}, \mathbf{x};\mathbf{b}, \mathbf{x}, t)\;.
   \label{eq:evo_hh}
\end{align}
Taking the moments
\begin{equation}
  \langle h_i(\mathbf{x},t)h_j(\mathbf{x'},t) \rangle= \int \textrm{d}\mathbf{b} b_{i}
  \int \textrm{d}\mathbf{b'} b'_{j}
  f^{\mathbf{h}\mathbf{h}}(\mathbf{b}, \mathbf{x};
  \mathbf{b'}, \mathbf{x'}, t)\;,
\end{equation}
of eq. (\ref{eq:evo_hh}) yields
\begin{align}
  \lefteqn{\frac{\partial}{\partial t}
  \langle h_i(\mathbf{x},t)h_j(\mathbf{x'},t)
  \rangle
  + \frac{\partial}{\partial x_{n}}
    \langle u_n(\mathbf{x},t) h_i(\mathbf{x},t)h_j(\mathbf{x'},t)\rangle
  + \frac{\partial}{\partial x'_{n}}
    \langle u_n(\mathbf{x'},t) h_i(\mathbf{x},t)h_j(\mathbf{x'},t)\rangle} \\ \nonumber
    &= \frac{\partial}{\partial x_{n}}
      \langle h_n(\mathbf{x},t) u_i(\mathbf{x},t)h_j(\mathbf{x'},t) \rangle
    + \frac{\partial}{\partial x'_{n}}
      \langle h_n(\mathbf{x'},t) h_i(\mathbf{x},t)u_j(\mathbf{x'},t)\rangle
      + \lambda  [\nabla_{\mathbf{x}}^2+\nabla_{\mathbf{x'}}^2]
      \langle h_i(\mathbf{x},t)h_j(\mathbf{x'},t)
      \rangle\;.
\end{align}
\section{Gaussian Approximation for the Joint Characteristic Functional of the MHD Equations}
\label{app:gauss}
In this appendix, we derive a cumulant expansion for the joint characteristic functional (\ref{eq:cumu_mhd}) and consider the implications of the vanishing of cumulants higher than order two.
In order to expand the joint characteristic functional (\ref{eq:cumu_mhd}) in powers of the fields $\boldsymbol{\alpha}(\mathbf{x})$
and $\boldsymbol{\beta}(\mathbf{x})$, it is convenient to introduce the six-dimensional fields $\mathbf{U}(\mathbf{x},t)=
\begin{pmatrix}\mathbf{u}(\mathbf{x},t)\\
  \mathbf{h}(\mathbf{x},t)\end{pmatrix}$ and
$\boldsymbol{\gamma}(\mathbf{x})=
\begin{pmatrix}\boldsymbol{\alpha}
  (\mathbf{x})\\
\boldsymbol{\beta}(\mathbf{x})\end{pmatrix}$. We can now expand the characteristic functional (\ref{eq:cumu_mhd}) in a power series
\begin{align}\nonumber
  \lefteqn{\phi[\boldsymbol{\gamma}(\mathbf{x}),t]
  = \left \langle e^{i \int \textrm{d}\mathbf{x} \boldsymbol{\gamma}(\mathbf{x})\cdot \mathbf{U}(\mathbf{x},t)}  \right \rangle}\\ \nonumber
  =&\phi[0] + \int \textrm{d} \mathbf{ x'}
 \left.\frac{\delta  \phi[\boldsymbol{\gamma} (\mathbf{ x}),t]}{\delta \gamma_I(\mathbf{x}')}
 \right |_{\boldsymbol{\gamma} (\mathbf{ x})=0}
 \gamma_I(\mathbf{x'})+\frac{1}{2!}\int \textrm{d} \mathbf{ x'}  \int  \textrm{d} \mathbf{ x''}   \left.\frac{\delta^2
 \phi[\boldsymbol{\gamma} (\mathbf{ x}),t]}{\delta \gamma_I(\mathbf{ x'})\delta \gamma_J(\mathbf{ x''})}
 \right |_{\boldsymbol{\gamma} (\mathbf{x})=0}
 \gamma_I(\mathbf{ x})   \gamma_J(\mathbf{ x''}) + \textrm{h.o.t.}\\ \nonumber
 =&1+ i \int \textrm{d} \mathbf{ x'}  \left \langle U_I(\mathbf{x'},t) \right \rangle\gamma_I(\mathbf{x'})
 + \frac{i^2}{2!} \int  \textrm{d} \mathbf{x'} \int  \textrm{d} \mathbf{x''}
  \left \langle U_I(\mathbf{ x'},t) U_J(\mathbf{x''},t) \right \rangle \gamma_I(\mathbf{x'})\gamma_J(\mathbf{x''})
 +\textrm{h.o.t.} \\
 =& 1+ \sum_{n=1}^{\infty} \int \textrm{d}\mathbf{ x}_1\ldots \int
 \textrm{d}\mathbf{ x}_n \frac{i^n}{n!} \gamma_{I_1}(\mathbf{ x}_1)\ldots\gamma_{I_n}(\mathbf{x}_n) C_{I_1\ldots I_n}(\mathbf{x}_1,
 \ldots,\mathbf{x}_n,t)\;,
 \label{eq:char_power_exp}
\end{align}
where capital indices indicate $I=1,\ldots,6$, and where we implied summation over equal indices. Moreover, we defined the correlation function
\begin{equation}
  C_{I_1\ldots I_n}(\mathbf{x}_1,
  \ldots,\mathbf{x}_n,t) = \left \langle U_{I_1}(\mathbf{x}_1,t) U_{I_2}(\mathbf{x}_2,t)\ldots U_{I_n}(\mathbf{x}_n,t) \right \rangle\;.
\end{equation}
On the other hand, we can cast (\ref{eq:char_power_exp}) in terms of the cumulants according to
\begin{equation}
  \phi[\boldsymbol{\gamma}(\mathbf{x}),t]
  = \exp \left( \sum_{n=1}^{\infty}\int \textrm{d}\mathbf{ x}_1\ldots \int
  \textrm{d}\mathbf{ x}_n \frac{i^n}{n!} \gamma_{I_1}(\mathbf{ x}_1)\ldots\gamma_{I_n}(\mathbf{x}_n)K_{I_1\ldots I_n}(\mathbf{x}_1,
  \ldots,\mathbf{x}_n,t)\right)\;,
\end{equation}
which suggests that the first four cumulants $K$ are related to the correlations functions $C$
according to
\begin{align}
  K_{I}=&C_{I}\;, \\
  K_{IJ}=&C_{IJ}-C_I C_J\;, \\
  \label{eq:three}
  K_{IJL}=&C_{IJL} -C_{IJ}C_{L} -C_{IL}C_{J}-C_{JL}C_{I}+2 C_{I}C_{J}C_{L} \;,\\ \nonumber
  K_{IJLM}=&C_{IJLM}-C_{IJ}C_{LM}-C_{IL}C_{JM}-C_{IM}C_{JL}-C_{IJL}C_M-C_{IJM}C_L  -C_{IML}C_J-C_{JLM}C_I \\ \nonumber
  ~&+2C_{IJ}C_L C_M+2C_{IL}C_J C_M +2 C_{IM}C_J C_L+2C_{ML}C_I C_J+2 C_{JL}C_I C_M+2C_{JM}C_I C_L\\
  ~&-6C_I C_J C_L C_M\;.
  \label{eq:four}
\end{align}
Under the assumption that there are no mean fields, i.e., $K_i=0$ and in neglecting higher order cumulants (\ref{eq:three}), (\ref{eq:four}), etc., we obtain
\begin{equation}
  \phi[\boldsymbol{\gamma}(\mathbf{x}),t]=
  e^{-\frac{1}{2}\int \textrm{d}\mathbf{x'} \int \textrm{d}\mathbf{x''} \gamma_I(\mathbf{x'}) K_{I\;J}(\mathbf{x'};\mathbf{x''},t) \gamma_J(\mathbf{x''})}
\end{equation}
Inserting the original fields $\mathbf{u}(\mathbf{x},t)$ and $\mathbf{h}(\mathbf{x},t)$ as well as the correlation functions (\ref{eq:cuu}-\ref{eq:chh}) thus yields
\begin{equation}
  \varphi[\boldsymbol{\alpha}(\mathbf{x}),\boldsymbol{\beta}(\mathbf{x}),t]
  = e^{-\frac{1}{2}\int \textrm{d}\mathbf{x'} \int \textrm{d}\mathbf{x''}  \left[\alpha_i(\mathbf{x'}) C^{\mathbf{u}\mathbf{u}}_{i\;j}(\mathbf{x'};\mathbf{x''},t) \alpha_j(\mathbf{x''})
  +2\alpha_i(\mathbf{x'},t)C^{\mathbf{u}\mathbf{h}}_{i\;j}(\mathbf{x'};\mathbf{x''},t) \beta_j(\mathbf{x''})
  +\beta_i(\mathbf{x'}) C^{\mathbf{h}\mathbf{h}}_{i\;j}
  (\mathbf{x'};\mathbf{x''},t) \beta_j(\mathbf{x''}) \right]}\;.
  \label{eq:gauss_func}
\end{equation}
In order to calculate the conditional expectation value
that enters the evolution equation for the single-point magnetic field, we consider the following relation
\begin{align}
  \left \langle u_i(\mathbf{x'},t) \delta(\mathbf{b}-\mathbf{h}(\mathbf{x},t)) \right\rangle
  = \frac{1}{(2\pi)^3}\int \textrm{d} \mathbf{w} e^{-i\mathbf{w}\cdot \mathbf{b}} \left. \frac{\delta \phi[\boldsymbol{\alpha}(\mathbf{x}),\boldsymbol{\beta}
  (\mathbf{x}),t]}{\delta (i \alpha_i(\mathbf{x'}))} \right|_{\boldsymbol{\beta}(\mathbf{x''})= \mathbf{w}\delta(\mathbf{x''}-\mathbf{x})}^{\boldsymbol{\alpha}(\mathbf{x''})=0}\;.
\end{align}
Inserting the Gaussian approximation of the characteristic functional (\ref{eq:gauss_func}) yields
\begin{align}\nonumber
   \left \langle u_i(\mathbf{x'},t) \delta(\mathbf{b}-\mathbf{h}(\mathbf{x},t)) \right\rangle
   = & \frac{i}{(2\pi)^3}\int \textrm{d} \mathbf{w} e^{-i\mathbf{w}\cdot \mathbf{b}}
 C^{\mathbf{u}\mathbf{h}}_{i\;j}(\mathbf{x'};\mathbf{x},t) w_j \left \langle e^{i\mathbf{w}\cdot \mathbf{h}(\mathbf{x},t)} \right \rangle \\ \nonumber
 =&-
 C^{\mathbf{u}\mathbf{h}}_{i\;j}(\mathbf{x'};\mathbf{x},t) \frac{\partial}{\partial b_j} \frac{1}{(2\pi)^3}\int \textrm{d} \mathbf{w} e^{-i\mathbf{w}\cdot \mathbf{b}}
 \left \langle e^{i\mathbf{w}\cdot \mathbf{h}(\mathbf{x},t)} \right \rangle\\
 =& -C^{\mathbf{u}\mathbf{h}}_{i\;j}(\mathbf{x'};\mathbf{x},t) \frac{\partial}{\partial b_j} f^{\mathbf{h}}(\mathbf{b},\mathbf{x},t)\;.
\end{align}
Accordingly, the conditional moment (\ref{eq:cond_u})
can be expressed as
\begin{equation}
  \left \langle u_i(\mathbf{x'},t)|\mathbf{b},\mathbf{x},t \right\rangle= C^{\mathbf{u}\mathbf{h}}_{i\;j}(\mathbf{x'};\mathbf{x},t) C^{\mathbf{h}\mathbf{h}}_{j\;k}(\mathbf{x};\mathbf{x},t)^{-1} b_k\;,
\end{equation}
where $C^{\mathbf{h}\mathbf{h}}_{j\;k}(\mathbf{x};\mathbf{x'},t)^{-1}$ denotes the inverse of the two-point tensor (\ref{eq:chh}).

\externalbibliography{yes}
\bibliography{lmn_mhd.bib}

\begin{thebibliography}{-------}
\providecommand{\natexlab}[1]{#1}

\bibitem[Chandrasekhar(1951)]{chandra:1951}
Chandrasekhar, S.
\newblock {The Invariant Theory of Isotropic Turbulence in
  Magneto-Hydrodynamics}.
\newblock {\em Proc. R. Soc. Lond. A} {\bf 1951}, {\em 204},~435--449.

\bibitem[Batchelor(1950)]{Batchelor1950}
Batchelor, G.K.
\newblock On the spontaneous magnetic field in a conducting liquid in turbulent
  motion.
\newblock {\em Proc. R. Soc. Lond. A} {\bf 1950}, {\em 201},~405--416.

\bibitem[Steenbeck \em{et~al.}(1966)Steenbeck, Krause, and R\"adler]{steenbeck}
Steenbeck, M.; Krause, F.; R\"adler, K.H.
\newblock Berechnung der mittleren Lorentz-Feldstärke für ein elektrisch
  leitendes Medium in turbulenter, durch Coriolis-Kräfte beeinflußter
  Bewegung.
\newblock {\em Z. Naturforsch} {\bf 1966}, {\em 21a},~369--76.

\bibitem[Parker(1996)]{Parker1996}
Parker, E.
\newblock S. Chandrasekhar and magnetohydrodynamics.
\newblock {\em J. Astrophys. Astron.} {\bf 1996}, {\em 17},~147--166.

\bibitem[R{\"a}dler(2007)]{Radler2007}
R{\"a}dler, K.H., Mean-Field Dynamo Theory: Early Ideas and Today's Problems.
\newblock In {\em Magnetohydrodynamics: Historical Evolution and Trends};
  Springer Netherlands: Dordrecht,  2007; pp. 55--72.

\bibitem[Galtier \em{et~al.}(2000)Galtier, Nazarenko, Newell, and
  Pouquet]{galtier_2000}
Galtier, S.; Nazarenko, S.V.; Newell, A.C.; Pouquet, A.
\newblock A weak turbulence theory for incompressible magnetohydrodynamics.
\newblock {\em J. Plasma Phys.} {\bf 2000}, {\em 63},~447–488.

\bibitem[Boldyrev(2006)]{Boldyrev2006}
Boldyrev, S.
\newblock {Spectrum of Magnetohydrodynamic Turbulence}.
\newblock {\em Phys. Rev. Lett.} {\bf 2006}, {\em 96},~115002.

\bibitem[Perez and Boldyrev(2010)]{Perez2010}
Perez, J.C.; Boldyrev, S.
\newblock Strong magnetohydrodynamic turbulence with cross helicity.
\newblock {\em Phys. Plasmas} {\bf 2010}, {\em 17},~055903.

\bibitem[Friedrich \em{et~al.}(2016)Friedrich, Homann, Sch{\"a}fer, and
  Grauer]{Friedrich2016}
Friedrich, J.; Homann, H.; Sch{\"a}fer, T.; Grauer, R.
\newblock Longitudinal and transverse structure functions in high
  Reynolds-number magneto-hydrodynamic turbulence.
\newblock {\em New J. Phys.} {\bf 2016}, {\em 18},~125008.

\bibitem[Hill(2001)]{hill:2001}
Hill, R.J.
\newblock {Equations relating structure functions of all orders}.
\newblock {\em J. Fluid Mech.} {\bf 2001}, {\em 434},~379.

\bibitem[Yakhot(2006)]{yakhot:2006}
Yakhot, V.
\newblock {Probability densities in strong turbulence}.
\newblock {\em Phys. D} {\bf 2006}, {\em 215},~166--174.

\bibitem[Grauer \em{et~al.}(2012)Grauer, Homann, and
  Pinton]{grauer-homann-pinton:2012}
Grauer, R.; Homann, H.; Pinton, J.F.
\newblock {Longitudinal and transverse structure functions in
  high-Reynolds-number turbulence}.
\newblock {\em New J. Phys.} {\bf 2012}, {\em 14},~63016.

\bibitem[Sorriso-Valvo \em{et~al.}(2007)Sorriso-Valvo, Marino, Carbone,
  Noullez, Lepreti, Veltri, Bruno, Bavassano, and
  Pietropaolo]{PhysRevLett.99.115001}
Sorriso-Valvo, L.; Marino, R.; Carbone, V.; Noullez, A.; Lepreti, F.; Veltri,
  P.; Bruno, R.; Bavassano, B.; Pietropaolo, E.
\newblock Observation of Inertial Energy Cascade in Interplanetary Space
  Plasma.
\newblock {\em Phys. Rev. Lett.} {\bf 2007}, {\em 99},~115001.

\bibitem[Fraternale \em{et~al.}(2019)Fraternale, Pogorelov, Richardson, and
  Tordella]{Fraternale_2019}
Fraternale, F.; Pogorelov, N.V.; Richardson, J.D.; Tordella, D.
\newblock Magnetic turbulence spectra and intermittency in the heliosheath and
  in the local interstellar medium.
\newblock {\em Astrophys. J.} {\bf 2019}, {\em 872},~40.

\bibitem[Bourgoin \em{et~al.}(2002)Bourgoin, Mari{\'e}, P{\'e}tr{\'e}lis,
  Gasquet, Guigon, Luciani, Moulin, Namer, Burguete, Chiffaudel,
  et~al.]{Bourgoin2002}
Bourgoin, M.; Mari{\'e}, L.; P{\'e}tr{\'e}lis, F.; Gasquet, C.; Guigon, A.;
  Luciani, J.B.; Moulin, M.; Namer, F.; Burguete, J.; Chiffaudel, A.; others.
\newblock Magnetohydrodynamics measurements in the von K{\'a}rm{\'a}n sodium
  experiment.
\newblock {\em Phys. Fluids} {\bf 2002}, {\em 14},~3046--3058.

\bibitem[Giesecke \em{et~al.}(2018)Giesecke, Vogt, Gundrum, and
  Stefani]{giesecke_2018}
Giesecke, A.; Vogt, T.; Gundrum, T.; Stefani, F.
\newblock Nonlinear Large Scale Flow in a Precessing Cylinder and Its Ability
  To Drive Dynamo Action.
\newblock {\em Phys. Rev. Lett.} {\bf 2018}, {\em 120},~024502.

\bibitem[Shew \em{et~al.}(2002)Shew, Sisan, and Lathrop]{shew2002mechanically}
Shew, W.; Sisan, D.; Lathrop, D.
\newblock Mechanically forced and thermally driven flows in liquid sodium.
\newblock {\em Magnetohydrodynamics} {\bf 2002}, {\em 38},~121.

\bibitem[Frisch(1995)]{frisch:1995}
Frisch, U.
\newblock {\em {Turbulence}}; Cambridge University Press,  1995.

\bibitem[Grauer and Marliani(1996)]{Grauer1996}
Grauer, R.; Marliani, C.
\newblock Analytical and numerical approaches to structure functions in
  magnetohydrodynamic turbulence.
\newblock {\em Phys. Scr.} {\bf 1996}, {\em 1996},~38.

\bibitem[Politano and Pouquet(1995)]{politano1995}
Politano, H.; Pouquet, A.
\newblock Model of intermittency in magnetohydrodynamic turbulence.
\newblock {\em Phys. Rev. E} {\bf 1995}, {\em 52},~636--641.
\newblock
  doi:{\changeurlcolor{black}\href{https://doi.org/10.1103/PhysRevE.52.636}{\detokenize{10.1103/PhysRevE.52.636}}}.

\bibitem[Chandrasekhar(2013)]{chandra:1961}
Chandrasekhar, S.
\newblock {\em {Hydrodynamic and hydromagnetic stability}}; Courier
  Corporation,  2013.

\bibitem[Friedrich \em{et~al.}(2018)Friedrich, Margazoglou, Biferale, and
  Grauer]{Friedrich2018}
Friedrich, J.; Margazoglou, G.; Biferale, L.; Grauer, R.
\newblock {Multiscale velocity correlations in turbulence and Burgers
  turbulence: Fusion rules, Markov processes in scale, and multifractal
  predictions}.
\newblock {\em Phys. Rev. E} {\bf 2018}, {\em 98},~023104.

\bibitem[Friedrich(2017)]{Friedrich2017}
Friedrich, J.
\newblock Closure of the Lundgren-Monin-Novikov hierarchy in turbulence via a
  Markov property of velocity increments in scale.
\newblock PhD thesis, Ruhr-University Bochum,  2017.

\bibitem[Lundgren(1967)]{Lundgren1967}
Lundgren, T.S.
\newblock Distribution functions in the statistical theory of turbulence.
\newblock {\em Phys. Fluids} {\bf 1967}, {\em 10},~969--975.

\bibitem[Monin(1967)]{Monin1967}
Monin, A.S.
\newblock {Equations of turbulent motion}.
\newblock {\em J. Appl. Math. Mech.} {\bf 1967}, {\em 31},~1057--1068.

\bibitem[Novikov(1968)]{Novikov1968}
Novikov, E.A.
\newblock {Kinetic Equations for a Vortex Field}.
\newblock {\em Sov. Phys. Dokl.} {\bf 1968}, {\em 12},~1006.

\bibitem[Friedrich \em{et~al.}(2012)Friedrich, Daitche, Kamps, L{\"{u}}lff,
  Vo{\ss}kuhle, and Wilczek]{Friedrich2012a}
Friedrich, R.; Daitche, A.; Kamps, O.; L{\"{u}}lff, J.; Vo{\ss}kuhle, M.;
  Wilczek, M.
\newblock {The Lundgren–Monin–Novikov hierarchy: Kinetic equations for
  turbulence}.
\newblock {\em Comptes Rendus Phys.} {\bf 2012}, {\em 13},~929--953.

\bibitem[Robertson(1940)]{Robertson1940}
Robertson, H.P.
\newblock {The invariant theory of isotropic turbulence}.
\newblock  Math. Proc. Cambridge Philos. Soc. Cambridge Univ Press,  1940,
  Vol.~36, pp. 209--223.

\bibitem[Keller and Friedman(1924)]{Keller1924}
Keller, L.V.; Friedman, A.A.
\newblock {Differentialgleichung f{\"{u}}r die turbulent Bewegung einer
  kompressiblen Fl{\"{u}}ssigkeit}.
\newblock  Proc. 1st Intern. Congr. Appl. Delft,  1924, pp. 395--405.

\bibitem[Ulinich and Lyubimov(1969)]{Ulinich1969a}
Ulinich, F.R.; Lyubimov, B.Y.
\newblock {The statistical theory of turbulence of an incompressible fluid at
  large Reynolds numbers}.
\newblock {\em Sov. J. Exp. Theor. Phys.} {\bf 1969}, {\em 28},~494.

\bibitem[Stanisic(2012)]{stanisic2012mathematical}
Stanisic, M.M.
\newblock {\em The mathematical theory of turbulence}; Springer Science \&
  Business Media,  2012.

\bibitem[Hopf(1952)]{Hopf1952}
Hopf, E.
\newblock Statistical hydromechanics and functional calculus.
\newblock {\em J. Rat. Mech. Anal.} {\bf 1952}, {\em 1},~87--123.

\bibitem[Grauer \em{et~al.}(1994)Grauer, Krug, and Marliani]{grauer1994scaling}
Grauer, R.; Krug, J.; Marliani, C.
\newblock Scaling of high-order structure functions in magnetohydrodynamic
  turbulence.
\newblock {\em Phys. Lett. A} {\bf 1994}, {\em 195},~335--338.

\bibitem[Friedrich \em{et~al.}(2012)Friedrich, Vo{\ss}kuhle, Kamps, and
  Wilczek]{Friedrich2012}
Friedrich, R.; Vo{\ss}kuhle, M.; Kamps, O.; Wilczek, M.
\newblock Two-point vorticity statistics in the inverse cascade of
  two-dimensional turbulence.
\newblock {\em Phys. Fluids} {\bf 2012}, {\em 24},~125101.

\bibitem[Wilczek and Meneveau(2014)]{Wilczek2014}
Wilczek, M.; Meneveau, C.
\newblock Pressure Hessian and viscous contributions to velocity gradient
  statistics based on Gaussian random fields.
\newblock {\em J. Fluid Mech.} {\bf 2014}, {\em 756},~191--225.

\bibitem[Chandrasekhar(1950)]{Chandrasekhar1950}
Chandrasekhar, S.
\newblock {The Theory of Axisymmetric Turbulence}.
\newblock {\em Philos. Trans. R. Soc. A} {\bf 1950}, {\em 242},~557--577.

\bibitem[Wilczek \em{et~al.}(2011)Wilczek, Daitche, and
  Friedrich]{Wilczek2011c}
Wilczek, M.; Daitche, A.; Friedrich, R.
\newblock {On the velocity distribution in homogeneous isotropic turbulence:
  correlations and deviations from Gaussianity}.
\newblock {\em J. Fluid Mech.} {\bf 2011}, {\em 676},~191--217.

\end{thebibliography}

\end{document}